\documentclass[sigconf, nonacm]{acmart}
\pdfoutput=1

\newcommand\vldbavailabilityurl{URL_TO_YOUR_ARTIFACTS}
\newcommand\vldbpagestyle{plain} 

\usepackage{balance}
\usepackage{subcaption}
\usepackage[linesnumbered, ruled, vlined]{algorithm2e}
\usepackage{xcolor}
\usepackage{xspace}
\usepackage{listings}

\usepackage{subfloat}
\usepackage{url}
\usepackage{soul}
\usepackage{times}
\sethlcolor{lightgray}
\usepackage[export]{adjustbox}
\usepackage{enumitem}
\usepackage{tabularx}
\usepackage{graphicx}
\graphicspath{{figures/}}
\usepackage{indentfirst}
\usepackage[normalem]{ulem}
\usepackage[inline]{aplcomments}
\newcommenter{yxy}{1,0,0}
\newcommenter{zhihan}{0.08627450980392157, 0.6274509803921569, 0.5215686274509804}
\newcommenter{todo}{0.0, 0, 1.0}
\newcommenter{mahesh}{0.7,0.4,0.0}
\newcommenter{phil}{0.6,0.3,0.0}
\newcommenter{xinyu}{0.0,0.8,0.8}
\newcommand{\edit}[1]{{\textcolor{black} {#1}}}
\newcommand{\newedit}[1]{{\textcolor{black} {#1}}}

\newcommand{\secref}[1]{Section~\ref{#1}}
\newcommand{\figref}[1]{Figure~\ref{#1}}
\newcommand{\tabref}[1]{Table~\ref{#1}}

\newcommand{\listref}[1]{Listing~\ref{#1}}
\newcommand{\lstref}[1]{Listing~\ref{#1}}
\newcommand{\algoref}[1]{Algorithm~\ref{#1}}

\newcommand*\ttvar[1]{\texttt{\expandafter\dottvar\detokenize{#1}\relax}}
\newcommand*\dottvar[1]{\ifx\relax#1\else
  \expandafter\ifx\string-#1\string-\allowbreak\else#1\fi
  \expandafter\dottvar\fi}
\let\oldnl\nl
\definecolor{comment-color}{rgb}{0.5,0.1,0.1}
\newcommand{\nonl}{\renewcommand{\nl}{\let\nl\oldnl}}
\newcommand{\codeComment}[1]{\textnormal{\color{comment-color}{\textit{\#
#1}}}\unskip}
\newcommand{\ignore}[1]{}

\newcommand{\logfunc}[0]{\textit{LogOnce()} }
\newcommand{\name}{Cornus\xspace}
\newcommand{\prepare}{\texttt{VOTE-YES}\xspace}
\newcommand{\commit}{\texttt{COMMIT}\xspace}
\newcommand{\abort}{\texttt{ABORT}\xspace}

\newcommand{\subsubsec}[1]{\vspace{.05 in}\noindent \textit{\textbf{#1}}. }

\begin{document}

\title{Cornus: Atomic Commit for a Cloud DBMS with \\ Storage Disaggregation (Extended Version)}

\newcommand{\footremember}[2]{%
   \footnote{#2}
    \newcounter{#1}
    \setcounter{#1}{\value{footnote}}%
}
\newcommand{\footrecall}[1]{%
    \footnotemark[\value{#1}]%
} 
\newcommand\vldbauthors{Zhihan Guo, Xinyu Zeng, Kan Wu, Wuh-Chwen Hwang, Ziwei Ren, Xiangyao Yu, Mahesh Balakrishnan, Philip A. Bernstein}

\author{
    Zhihan Guo, Xinyu Zeng, Kan Wu, Wuh-Chwen Hwang, Ziwei Ren, \\ Xiangyao Yu, Mahesh Balakrishnan$^{\dagger}$, Philip A. Bernstein$^{\ddagger}$
}
\affiliation{%
  \institution{University of Wisconsin-Madison,\ \ \  $^{\dagger}$Confluent, Inc., \ \ \  $^{\ddagger}$Microsoft Research}
  \country{}
}
\email{{zhihan, xzeng, kanwu, wuh-chwen, ziwei, yxy}@cs.wisc.edu}
\email{mbalakrishnan@confluent.io,  
philbe@microsoft.com}

\renewcommand{\shortauthors}{Guo, Zeng, Wu, Hwang, Ren, Yu, Balakrishnan, Bernstein et al.}

\begin{abstract}
\noindent \textit{Two-phase commit} (2PC) is widely used in distributed databases to ensure  atomicity of distributed transactions. Conventional 2PC was originally designed for the shared-nothing architecture and has two limitations: \textit{long latency} due to two eager log writes on the critical path, and \textit{blocking} of progress when a coordinator fails.

Modern cloud-native databases are moving to a storage disaggregation architecture where storage is a shared highly-available service. 
Our key observation is that disaggregated storage enables protocol innovations that can address both the long-latency and blocking problems. 
We develop \name, an optimized 2PC protocol to achieve this goal. 
The only extra functionality \name requires is an atomic compare-and-swap capability in the storage layer, which many existing storage services already support.
We present \name in detail with proofs and show how it addresses the two limitations. We also deploy it on real storage services including Azure Blob Storage and Redis. Empirical evaluations show that \name can achieve up to 1.9$\times$ latency reduction over conventional 2PC. 
\end{abstract}

\maketitle

\pagestyle{\vldbpagestyle}


\begin{figure}[t]
    \begin{subfigure}[t]{.47\linewidth}%
    	\center
        \includegraphics[width=\linewidth]{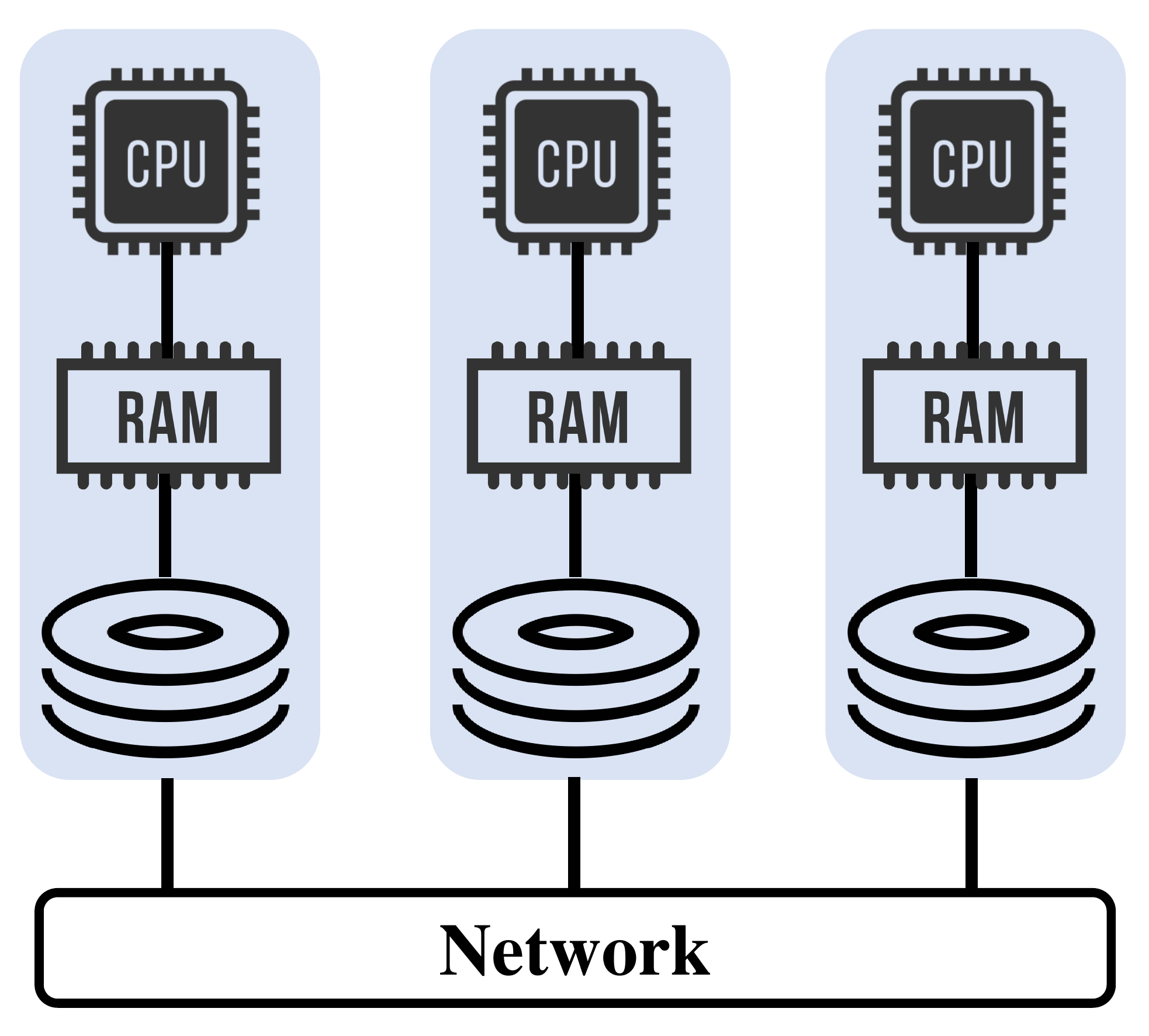}
        \caption{Shared-nothing}\label{fig:shared-nothing}
    \end{subfigure}%
    \hfill
    \begin{subfigure}[t]{.47\linewidth}%
    	\center
        \includegraphics[width=\linewidth]{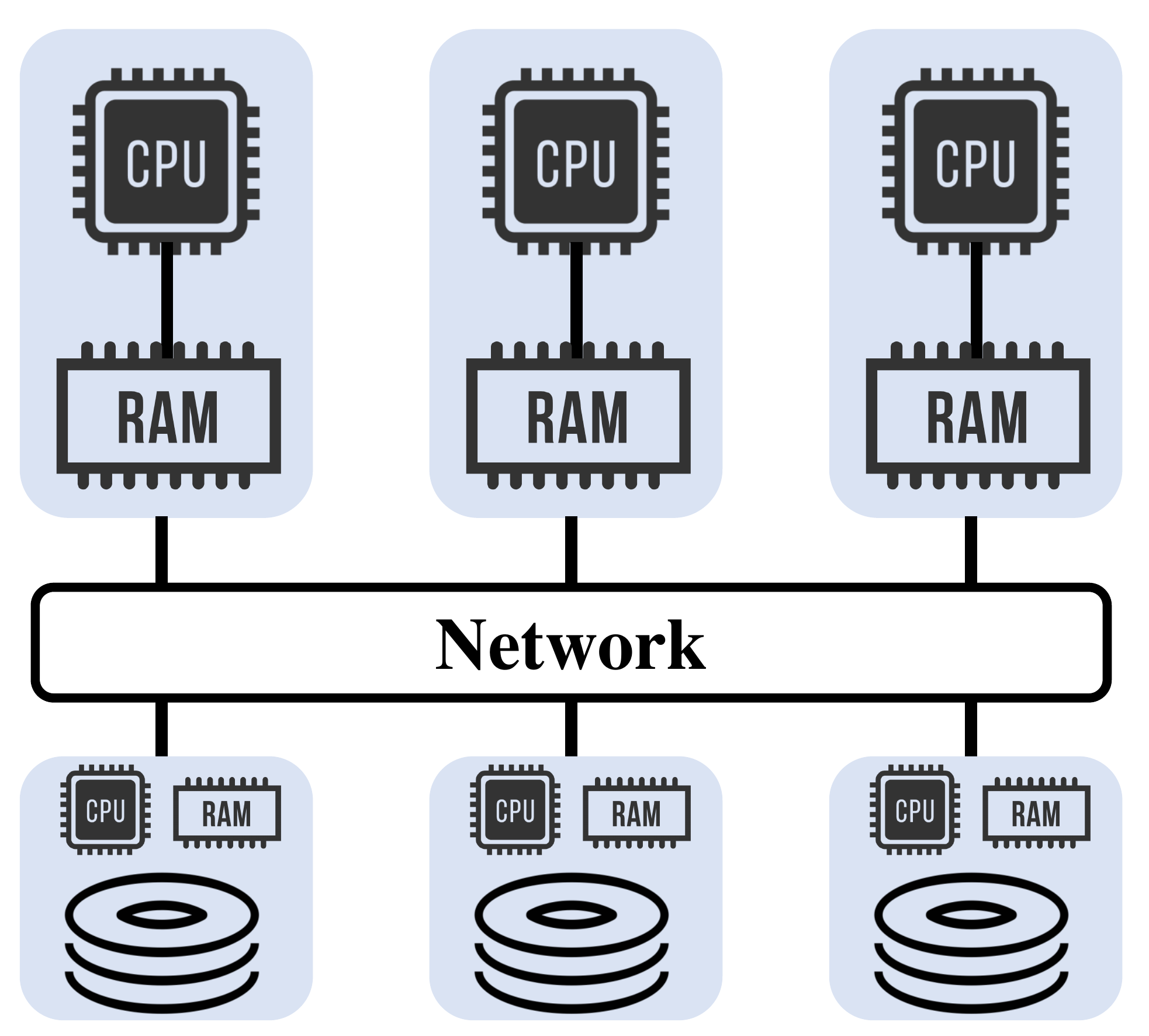}
        \caption{Storage-disaggregation}\label{fig:disaggregation}    
    \end{subfigure}%
    \caption{Shared-Nothing vs. Storage-Disaggregation. }
    \vspace{-.1in}
    \label{fig:arch_compare}
\end{figure}

\begin{figure*}[t!]
 	\begin{subfigure}[t]{.48\linewidth}%
 	\center
 	\includegraphics[width=\linewidth]{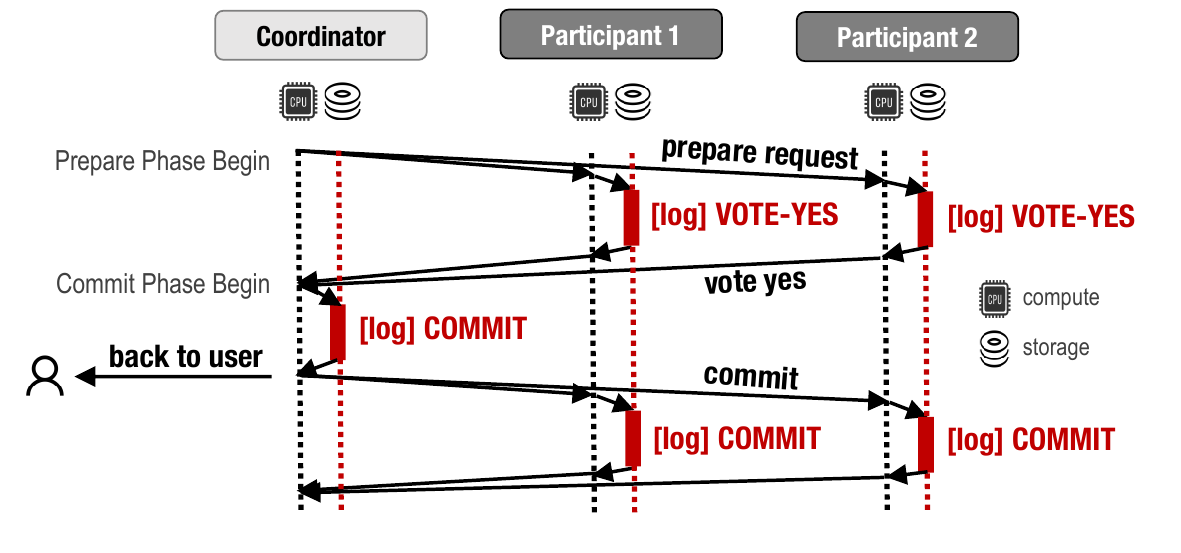}
 	\caption{2PC with no failure.}
 	\label{fig:2pc_commit}
 	\end{subfigure}
 	\hfill
 	\begin{subfigure}[t]{.48\linewidth}%
 	\center
 	\includegraphics[width=\linewidth]{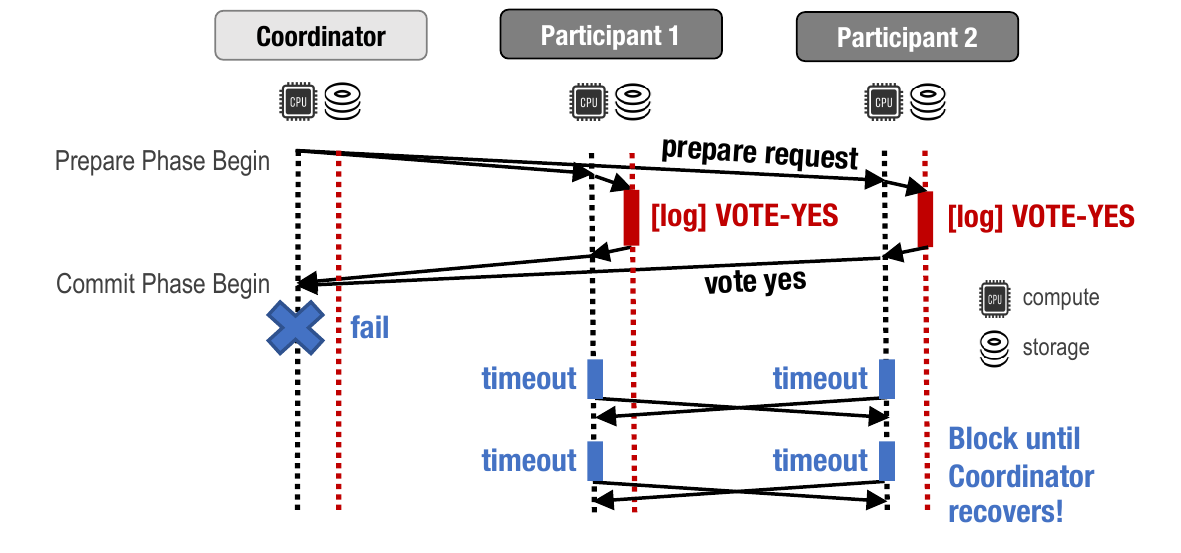}
 	\caption{2PC with coordinator failure (cooperative termination protocol). }
 	\label{fig:2pc_block}
 	\end{subfigure}
 	\caption{Illustration of Two-Phase Commit (2PC) --- \normalfont{The lifecycle of a committing transaction (a) and a scenario of coordinator failure (b).}}
 	\label{fig:2pc}
\end{figure*}

\section{Introduction} \label{sec:intro}
Databases are migrating to the cloud because of desirable features such as elasticity, high availability, and cost competitiveness.
Modern cloud-native databases feature a \textit{storage-disaggregation} architecture where the storage is decoupled from computation as a standalone service as shown in \figref{fig:disaggregation}. 
This architecture allows independent scaling and billing of computation and storage, which can improve resource utilization, reduce operational cost, and enable flexible cloud deployment with heterogeneous configurations.
Many cloud-native database systems adopt such an architecture for both OLTP~\cite{verbitski2017amazon, foundationdb, malkhi2013spanner, socrates} and OLAP~\cite{spectrum, athena, presto, hive, sparksql, snowflake}. Nowadays, as storage services offer essential functions such as fault tolerance, scalability, and security at low-cost, systems start to layer their designs on the existing disaggregated storage services~\cite{armburst2020deltalake,brantner2008building}.


This paper focuses on efficient deployment of the two-phase commit protocol on existing 
storage services. 
Two-phase commit (2PC) is the most widely used atomic commit protocol, which ensures that distributed transactions commit in either all or none of the involved data partitions. 
2PC was originally designed for the \textit{shared-nothing} architecture and suffers from two major problems.
The first is \textit{long latency}: 2PC requires two round-trip network messages and associated logging operations. Previous work has demonstrated that the majority of a transaction’s execution time can be attributed to 2PC
~\cite{yu2018sundial, dragojevic15, harding2017, al1995two, abdallah1998one, samaras1993two, mohan1986transaction}. 
The second problem is \textit{blocking}~\cite{skeen1981nonblocking, babaoglu1993understanding, bernstein1987concurrency}.
Blocking occurs if a coordinator crashes before notifying participants of the final decision. 
These two problems greatly limit the performance of 2PC, especially in a storage disaggregation architecture 

Various techniques have been proposed to address these two problems with 2PC. Some proposed optimizations target the shared-nothing architecture and do not solve both problems simultaneously.
These protocols either reduce latency by making strong assumptions about the workload and/or system that are not always practical for disaggregated storage~\cite{stamos1990low, stamos1993coordinator, al1995two, lee2002single, abdallah1997non,abdallah1998one,kung1981optimistic, bernstein1987concurrency}, or they mitigate the blocking problem by adding an extra phase and prolong latency~\cite{skeen1981nonblocking, gupta2018easycommit, babaoglu1993understanding}. Another line of research addresses both problems through customizing the storage. Examples include Paxos Commit~\cite{gray2006},  TAPIR~\cite{zhang2018building}, MDCC~\cite{kraska2013mdcc}, and parallel commit in CockroachDB~\cite{taft2020cockroachdb}. 
Existing solutions, however, are not applicable to general storage services because they require customized storage designs that perform
conflict detection between transactions~\cite{kraska2013mdcc, zhang2018building, cockroachdb, taft2020cockroachdb} and/or need specific replication protocols~\cite{gray2006, kraska2013mdcc, zhang2018building}. 
Therefore, they cannot be readily applied to most existing storage services.

\newedit{
In this paper, we aim to maximize the flexibility brought by disaggregation without requiring customized APIs for the storage service. Therefore, a database can adopt existing highly optimized storage services and thereby avoid the expense of developing a new one, and can also allow the storage to adopt new mechanisms (e.g., new replication protocols) independently. 
We aim to answer the following research question: 
\textit{What is the minimal requirement from the storage layer to enable 2PC optimizations addressing high latency and blocking?}
Our answer is that the only requirement is the ability to provide \emph{log-once} functionality, which ensures that for each transaction, only one update of its state in the log is allowed. 
We show that log-once semantics can be achieved with a simple \emph{compare-and-swap-like API}, which is supported by almost every storage service today, including Redis~\cite{redis}, Microsoft Azure Storage~\cite{calder2011windows}, Amazon Dynamo~\cite{decandia2007dynamo}, and Google BigTable~\cite{chang2008bigtable}. 
}

We introduce \textbf{\name}, an optimized 2PC protocol specifically designed for the storage disaggregation architecture in cloud-native databases. 
\name makes two major changes to conventional 2PC.
First, it eliminates decision logging by the coordinator, which significantly reduces the latency of 2PC. A transaction is committed as soon as all participants have written \prepare in their log in response to prepare requests in phase one. This optimization is feasible because highly-available disaggregated storage ensures that after log is written, it will not be lost. 
Second, \name uses a \logfunc API based on compare-and-swap functionality to address the blocking problem. 
The API allows multiple participants to read and update the same log file and ensures that only the first log append for a transaction can update this transaction's state. This feature allows any participant that is uncertain about the transaction's outcome to force an abort by writing a vote to an unresponsive participant's log.

We study the deployment of \name over commercial cloud storage services (e.g., Azure Table, Azure Blob, Redis, Amazon S3, etc.) and discuss its implications for performance and privacy. We implement \name in an open-source distributed DBMS, Sundial~\cite{yu2018sundial}, and deploy it on Azure Blob~\cite{calder2011windows} and Redis~\cite{redis}. 

In summary, this paper makes the following contributions:
\begin{itemize}
	\item We revisit the design of 2PC in the context of a storage-dsiaggregation architecture.
	\item We develop \name, an optimized 2PC protocol for the storage-disaggregation architecture. It reduces transaction latency during normal processing and alleviates the blocking problem. 
	\item We prove the correctness of \name by showing it satisfies all the required properties of an atomic commit protocol. 
	\item We deploy \name on practical storage services, Redis~\cite{redis} and Azure Blob~\cite{calder2011windows}, and show up to a 1.9$\times$ speedup in latency. 
\end{itemize}

The rest of the paper is organized as follows: Section 2 provides background about two-phase commit and the architectural changes brought by storage disaggregation. Section 3 describes \name, including pseudocode, optimizations for read-only transactions, and analyses of failure cases. Section 4 discusses the deployment of \name on practical storage services and Section 5 evaluates it. Section 6 discusses related work and Section 7 is the conclusion. 

\section{Background and Motivation} \label{sec:background}

\subsection{Two Phase Commit (2PC)}

\newedit{In a partitioned database, each partition has a corresponding process called a \emph{resource manager} that handles reads and writes to the partition. 
2PC is an atomic commit protocol that ensures a transaction involving multiple partitions either commits everywhere or aborts everywhere. 2PC contains a \emph{prepare phase} and a \emph{commit phase}, shown in \figref{fig:2pc}. }
\newedit{The process initiating 2PC is called the \emph{coordinator}. }
Resource managers that took part in the distributed transaction are called \emph{participants}. 

%
%

\figref{fig:2pc_commit} shows the logging events and network messages when all participating resource managers agree to commit the transaction and no failure occurs. If the coordinator fails before sending the decision to all participants, as shown in \figref{fig:2pc_block}, some participants may not know the decision. \newedit{A participant that 
voted yes and then times out waiting for the coordinator's decision
will initiate a \emph{termination protocol}. In a \emph{naive termination protocol}, the participant needs to wait until the coordinator recovers. If the coordinator includes a list of participants in the prepare request, the uncertain participants 
can run a \emph{cooperative termination protocol}. In this case, an uncertain participant contacts other participants to learn the decision. It repeats this process until at least one participant knows the decision as shown in \figref{fig:2pc_block}. 
}
In 2PC, the coordinator’s decision log record serves as the ground truth of the commit/abort decision --- the final outcome of the transaction relies on the success of logging this record. The conventional 2PC has the following two limitations.

\vspace{.05in}
\noindent\textbf{Limitation 1: Latency of Two Phases}

In the standard 2PC protocol, the transaction caller experiences an average latency of \textit{one network round-trip} and \textit{two logging operations}, as shown in \figref{fig:2pc_commit}. Such a delay directly affects the transaction response time that an end-user will experience.

\vspace{.05in}
\noindent\textbf{Limitation 2: Blocking Problem}

In 2PC, a participant learns the decision of a transaction either directly from the coordinator or indirectly from other participants. In an unfortunate corner case shown in~\figref{fig:2pc_block} where the coordinator fails before sending any notification, no participant can make or learn the decision.
\newedit{While the decision is uncertain, any new transaction that conflicts with the current uncertain transaction cannot commit, since it depends on the outcome of the uncertain transaction.
This is the well known blocking problem which causes data to be inaccessible due to the failure of another resource manager not holding the data, limiting the performance and data availability of 2PC. }

Many optimizations have been developed to improve the two limitations above in a shared-nothing architecture, including eliminating the prepare phase, at the cost of making extra system assumptions~\cite{stamos1990low, stamos1993coordinator, al1995two, lee2002single, abdallah1997non}, or eliminating blocking, at the cost of a third phase~\cite{skeen1981nonblocking, gupta2018easycommit, babaoglu1993understanding}. 
In Section~\ref{sec:related_work}, we describe these protocols in more detail and explain their relationship with \name.

\subsection{2PC in Storage-Disaggregation Architecture} \label{ssec:storage_disag}

Modern cloud-native DBMSs use a storage disaggregation architecture where storage and computation are separate services connected by the data center network (\figref{fig:disaggregation}), in contrast to shared-nothing systems with direct-attached storage (\figref{fig:shared-nothing}). Many cloud-native OLTP databases adopt this architecture including Amazon Aurora~\cite{verbitski2017amazon}, Apple FoundationDB~\cite{zhou2021foundationdb}, CockroachDB~\cite{cockroachdb}, Google Spanner~\cite{malkhi2013spanner}, and Microsoft SQL Server~\cite{socrates}. 
The disaggregation architecture allows the computation and storage layers to scale and be developed independently. It offers better elasticity, flexibility, and resource utilization, and separates the concerns to achieve easier management and lower operational cost. 

Naively deploying 2PC over a storage-disaggregation architecture increases the logging latency since the storage service is at the other side of the network. 
Previous research has tried to leverage the architectural features of storage disaggregation to optimize classic 2PC for better performance. 
Paxos Commit~\cite{gray2006} was the first to develop a theoretical framework. In particular, the protocol contains four key ideas: (1) A transaction's fate is no longer decided by the decision logging of the coordinator, but by the votes logged by all participants including the coordinator. 
(2) Paxos acceptors can pre-prepare to save the first phase in Paxos.
(3) Participants can directly propose the prepared message to replicas (i.e., acceptors), thereby skipping the Paxos leader to save one message delay. 
(4) The coordinator is also the Paxos leader of all Paxos instances. The coordinator can learn the decision directly from acceptors to save two more message delays.

Paxos Commit and its followup protocols~\cite{cockroachdb, kraska2013mdcc,mahmoud2013low, zhang2018building, yan2018carousel, fan2019ocean} co-design 2PC with the underlying consensus protocol in the storage service. While they are non-blocking and achieve low latency, they require significant customization of the storage layer. Thus, they cannot be readily deployed in existing storage services. 
In this paper, we aim to solve the long latency and blocking problems \textit{without} customizing the storage service. We aim to develop a protocol that largely retains the performance advantage of Paxos Commit-like optimizations while being deployable in existing storage services.

\section{\name} \label{sec:protocol}

\name is a non-blocking, low-latency 2PC variant that has minimal requirements on the disaggregated storage service. The only new storage-layer function needed by \name is the logging procedure \textit{LogOnce()}, which can be implemented using compare-and-swap --- a capability 
supported in many cloud-native storage services. This section presents the \name protocol in detail. 

After presenting the big picture in \secref{ssec:design}, we describe the APIs and protocols in Sections~\ref{ssec:storage_api} and \ref{ssec:protocol} respectively. \secref{ssec:failure} describes how \name handles failures and recovery. 
\secref{ssec:proof} proves the correctness of \name.  
\secref{ssec:opt} discusses read-only transactions and \secref{ssec:further-opt} discusses further optimizations. 

\begin{figure}[t]
    \includegraphics[width=\linewidth]{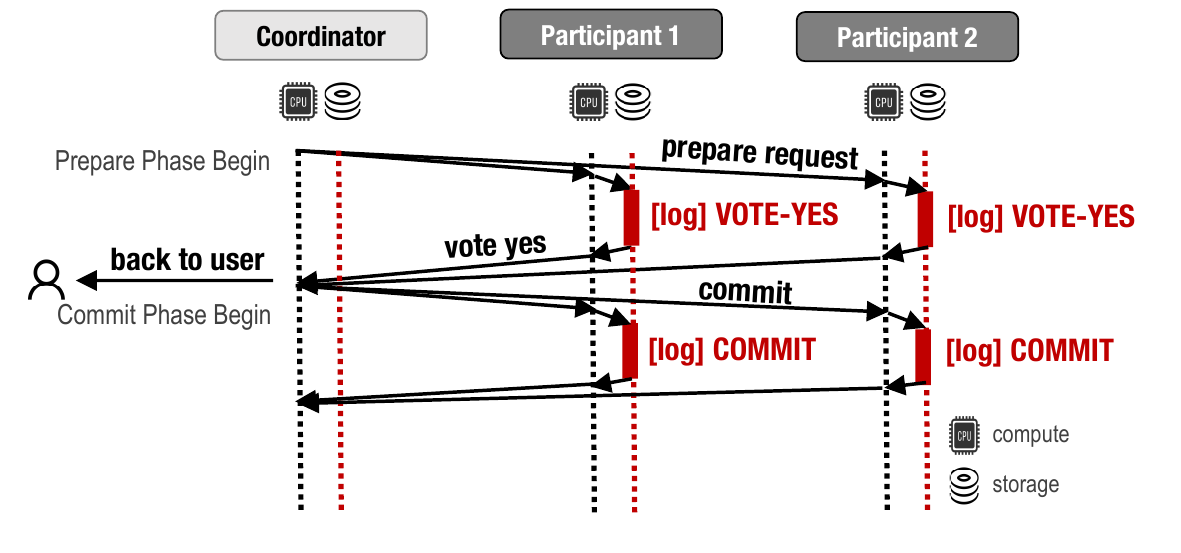}
    \caption{Illustration of \name.}
    \label{fig:1pc_commit}
\end{figure}

\subsection{Design Overview} \label{ssec:design}

Disaggregated storage enables the optimizations in Cornus as it has the following features that a conventional shared-nothing architecture does not provide: 

\noindent\textbf{Feature \#1:} A disaggregated storage service has \textit{built-in data replication to support  high availability}.

\noindent\textbf{Feature \#2:} A disaggregated storage service can be \textit{accessed by all the participants}. 

\noindent\textbf{Feature \#3:} A disaggregated storage service can support \textit{limited computation tasks beyond basic reads and writes}. 

Below we explain why these new features of storage disaggregation can enable optimizations that reduce commit latency and avoid blocking
at the same time.

\vspace{.05in}
\noindent\textbf{Latency reduction}. 
In conventional 2PC, the final outcome of a transaction is determined by the coordinator's decision log. In \name, we follow the first idea in Paxos Commit (cf. Section~\ref{ssec:storage_disag}) and change the criterion to be the \textit{collective votes in all participants logs}; namely, a transaction commits if and only if all participants have logged \prepare. 
\figref{fig:1pc_commit} shows the procedure of a committing transaction in \name. Compared to conventional 2PC as shown in \figref{fig:2pc_commit}, the coordinator no longer needs to log the decision, which eliminates this source of latency.

Features \#1 and \#2 are critical in achieving this latency reduction. 
They guarantee that once a vote is written to storage, all resource managers can access it most of the time. In Section~\ref{ssec:protocol}, we show how \name ensures correctness by leveraging these features. 

\vspace{.05in}
\noindent\textbf{Non-blocking}. 
Conventional 2PC can block when a coordinator fails and its decision log cannot be accessed. 
\name avoids blocking because the collective decision logs are written to the storage service (not the coordinator's local disk) and thus are highly available (Feature \#1).
If a participant is uncertain about the transaction's outcome, it can read the votes of all participants to learn that outcome. The uncertain participant can even write to a log on behalf of an unresponsive participant to enforce the final decision (Feature \#2). 
Here we assume that each data partition keeps one log. 
A single-partition transaction writes to the log in the corresponding data partition; a distributed transaction writes to logs of all corresponding partitions that it accesses. These behaviors are identical to 2PC.

If multiple participants try to resolve the outcome of an unresponsive participant by writing to the same log, they can create a data race. 
To avoid this, we leverage Feature \#3 and introduce a new API in the storage service called \textit{LogOnce()}. It guarantees that a transaction's state in the log is write-once ---
after the first update of the transaction's state, future updates to the state have no effect.
\logfunc is the only extra storage-layer API required by \name and is powerful enough to ensure correctness and avoid blocking. 
We will explain the \logfunc API in more detail in \secref{ssec:storage_api} and its implementation using commercial storage services in \secref{sec:deployment}.
We note that \name may still block due to a failure of the underlying storage system. However, in any storage-disaggregation database, the system would be blocked if the underlying storage becomes unavailable, since the database is not able to access any persistent data. 
Moreover, modern cloud storage services are typically highly available. 
For example, Azure Blob Storage has “11 nines” durability with locally redundant storage and “12 nines” with zone-redundant storage~\cite{azure-availability}.

\subsection{Cornus APIs}\label{ssec:storage_api}

We first describe our RPC notation for remote procedure calls and then introduce the logging APIs used in \name. 

\vspace{.05 in}
\noindent\textit{\textbf{Remote Procedure Calls (RPC)}}

We model communication between participants as RPCs. 
An RPC can be either \textit{synchronous} or \textit{asynchronous}. 
A synchronous call blocks until the callee returns.
An asynchronous call allows the caller to continue executing until it explicitly waits for the response. 

We represent an RPC using the following notation: 

\vspace{-.08in}
\[ \textbf{\textit{RPC}$_{\textit{sync/async}}^n$::\textit{FuncName()}} \]

\noindent where the subscript can be \textit{sync} or \textit{async} for synchronous and asynchronous RPCs respectively. The superscript $n$
denotes the destination. 
\textit{FuncName()} is the function to be called on the remote site and can take arbitrary arguments. 
In this paper, we consider the following two RPC functions:

\vspace{.05 in}
\noindent \textit{\textbf{Log(txn, type)}}

The \textit{Log(txn, type)} function simply appends a log record of a certain \textit{type} to the end of transaction \textit{txn}'s log. It is used in both the conventional 2PC protocol and \name.

\vspace{.05 in}
\noindent \textit{\textbf{LogOnce(txn, type)}}

In \name, we introduce \textit{LogOnce(txn, type)} to guarantee that a transaction's state can be written at most once.
It atomically checks if a log record already exists for \textit{txn}, and if not assigns the value in \textit{type} to the record, namely \prepare, \commit, or \abort.
It returns 
the state of \textit{txn} after performing the atomic operation, which is either \textit{type} or the existing state, depending on whether the operation updated the state. 
This function is used in \name but not in conventional 2PC.

\subsection{Cornus Protocol} \label{ssec:protocol}

\begin{algorithm}[t!]
	\small 
    \SetKwProg{myfun}{Function}{}{}
    \nonl\texttt{\\}
       
    \myfun{Coordinator::StartCornus(txn)}{ 
    	\For{p in txn.participants} {
    		\textbf{send} \texttt{VOTE-REQ} to $p$ asynchronously \\
    	}
    	\textbf{wait for }\textit{all responses from participants} \\
    	\Indp 
    	      \textbf{on receiving \texttt{ABORT}} \textit{decision $\leftarrow$ \texttt{ABORT}} \\              
    	      \textbf{on receiving all responses} \textit{decision $\leftarrow$ \texttt{COMMIT}} \\
              \textbf{on timeout } \hl{\textit{decision $\leftarrow$ TerminationProtocol(txn)}} \\
        \Indm
    	\hl{\textit{reply decision to the txn caller}} \\
    	\For{p in txn.participants}{
            \textbf{send} \textit{decision} to p asynchronously \\
    	}
    }
    \nonl\texttt{\\}
    
    \myfun{Participant::StartCornus(txn)}{
    	\textbf{wait for} \textit{\texttt{VOTE-REQ}} \textbf{from} \textit{coordinator }\\
    		\Indp \textbf{on timeout} 
            \textit{RPC$_{\textit{sync}}^{\textit{ local log}}$::Log(\texttt{ABORT})}
    		\Return\\
    	\Indm
    	\uIf{participant votes yes for txn} {
    		\textit{resp $\leftarrow$ RPC$_{\textit{sync}}^{\textit{ local log}}$::\hl{LogOnce}(\texttt{VOTE-YES})}\\
    		\uIf{\hl{resp is \texttt{\texttt{ABORT}}}} {
    			\nonl \codeComment{Another participant has logged \texttt{ABORT} for it} \\
    			\hl{\textbf{reply } \textit{\texttt{ABORT} to coordinator}}
    		}
    		\Else {
    			\textbf{reply } \textit{\texttt{VOTE-YES} to coordinator}  \\
    			\textbf{wait for} \textit{
                decision} \textbf{from} \textit{coordinator} \\
    				\Indp \textbf{on timeout }\textit{decision $\leftarrow$ TerminationProtocol(txn)} \\
    				\Indm
    				\textit{RPC$_{\textit{sync}}^{\textit{ local log}}$::Log(decision)}
    		}
    	}
    	\Else {
    		\textit{RPC$_{\textit{async}}^{\textit{ local log}}$::Log(\texttt{ABORT})}\\
    		\textbf{reply } \textit{\texttt{ABORT} to coordinator} \\
    	}
    }
    \nonl\texttt{\\}
    
    \myfun{\hl{TerminationProtocol(txn)}}{
    	\For{every paticipant p other than self} {
            \textit{{RPC$_{\textit{async}}^{\textit{ p.log}}$::LogOnce(\texttt{ABORT})}}\\
    	}
    	\textbf{wait for responses} \\
    	\Indp 
              \textbf{on receiving \texttt{ABORT}} decision $\leftarrow$ \texttt{ABORT}\\
              \textbf{on receiving \texttt{COMMIT}} decision $\leftarrow$ \texttt{COMMIT}\\              
              \textbf{on receiving all responses} decision $\leftarrow$ \texttt{COMMIT}\\              
              \textbf{on timeout} retry from the beginning \\
    	\Indm
    	\Return \textit{decision}
    }
    \caption{\textbf{API of Resource Managers in \name} --- Assuming a committing transaction. Differences between Cornus and 2PC are highlighted in \hl{gray}. }
    \label{alg:protocol}

\end{algorithm}

The pseudocode for \name is shown in~\algoref{alg:protocol}. 
We use a gray background to highlight the key changes in \name compared to standard 2PC \newedit{with cooperative termination protocol introduced in Section 2.1, where a coordinator will send out each prepare request along with a list of coordinator's and participants' addresses}. In the following, we explain the pseudocode for the coordinator, the participant, and the termination protocol. 

\vspace{.05 in}
\noindent\textit{\textbf{Coordinator::StartCornus(txn)}}

After a transaction \textit{txn} finishes the execution phase, the coordinator calls \textit{StartCornus(txn)} to start the atomic commit protocol.
The coordinator sends out vote requests along with a list of all participants involved in the transaction to all participants (lines 2--3). 
Then the coordinator waits for responses from all participants (line 4). 
If it receives an \abort, the transaction reaches an abort decision (line 5). 
If it receives \prepare from all participants (i.e., none of them is \abort), the transaction reaches a commit decision (line 6).
If it times out, it invokes the termination protocol to finalize a decision (line 7). 
The latter is unlike 2PC, which unilaterally aborts the transaction without running the termination protocol. 

Once the decision is reached, it can be returned to the transaction caller immediately without logging the decision. It is a key difference between \name and 2PC; the latter would reply to the caller only after the decision is written to stable storage. This optimization reduces the caller-observed latency by the duration of one logging operation. Finally, the coordinator 
broadcasts the decision to all participants asynchronously (lines 9--10).

\begin{figure*}[ht]
    \begin{subfigure}[b]{.5\linewidth}%
        \includegraphics[width=.95\linewidth]{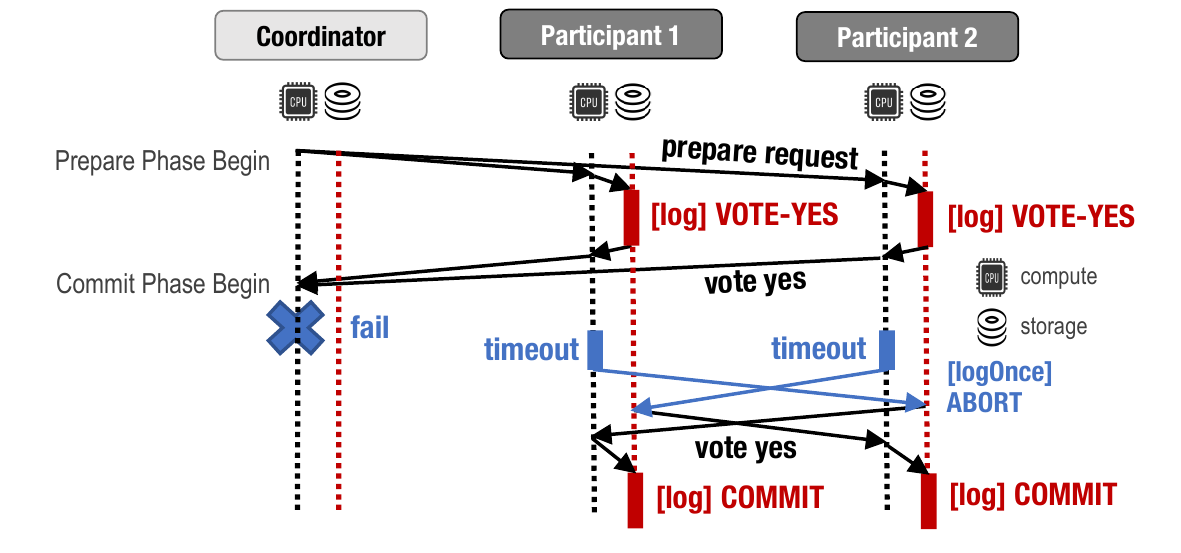}
        \caption{Coordinator fails before sending decision}\label{fig:1pc_block}
    \end{subfigure}%
    \begin{subfigure}[b]{.5\linewidth}%
        \includegraphics[width=.95\linewidth]{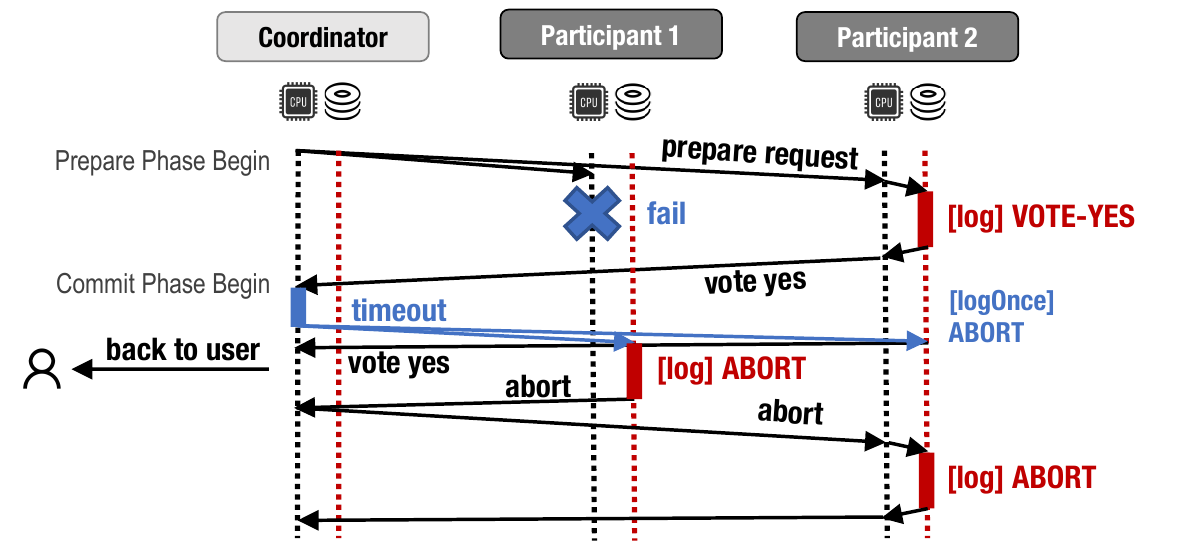}
        \caption{Participant fails before logging vote}\label{fig:1pc_abort}
    \end{subfigure}%
    \caption{\name under Failures --- \normalfont{The behavior of \name under two failures scenarios.
    }}
    \vspace{-.15in}
    \label{fig:1pc_failure}
\end{figure*}

\vspace{.05 in}
\noindent\textit{\textbf{Participant::StartCornus(txn)}}

The participant logic is very similar for \name and 2PC. 
A participant waits for a \texttt{VOTE-REQ} message from the coordinator (line 12). 
If it times out, it can unilaterally abort the transaction (line 13).

After receiving a \texttt{VOTE-REQ}, a participant votes \prepare or \texttt{VOTE-NO} based on its local state of the transaction. For a \texttt{VOTE-NO}, it logs an \abort record and replies to the coordinator (lines 24--25). 
This logging operation can be asynchronous following the \textit{presumed abort} optimization in conventional 2PC~\cite{mohan1986transaction}.

For a \prepare, the participant logs the record using \logfunc (line 15). There are two possible outcomes. If the function returns \abort, then another participant already aborted the transaction on behalf of this participant through the termination protocol. In this case, the participant aborts the transaction and returns \abort to the coordinator (lines 16--17). 
Otherwise the function returns \prepare, in which case the participant returns \prepare to the coordinator (lines 19) and starts to wait for the coordinator's decision message (line 20). 
If it times out, it executes the termination protocol (lines 21). 
Otherwise, the decision received from the coordinator is written to the local log (line 22).

\vspace{.05 in}
\noindent \textit{\textbf{TerminationProtocol(txn)}}

In both 2PC and \name, a participant executes the termination protocol when it times out while waiting for a message and cannot unilaterally abort the transaction. 
In 2PC, the participant running the cooperative termination protocol contacts all the other participants for the outcome of the transaction. 
If any participant returns the outcome, the uncertainty is resolved. Otherwise, it will block until the failure is recovered. 

\name avoids this problem. Instead of contacting the resource managers, the participant running the termination protocol tries to log an \abort record in each participant's log 
using the function \logfunc (lines 27--28). 
\newedit{If the remote storage has not received any log record for the transaction yet, the \abort record will be logged and returned (line 30). 
If the log already received a decision log record (i.e., \commit or \abort) for this transaction, the function will return that decision (line 30--31). }
Another case is that the \logfunc returns a \prepare record. If the current participant receives this response from all the logs, 
it commits the transaction (line 32). 
If the current participant experiences a timeout again, it retries the termination protocol (line 33). 

The only case where \name blocks is when it cannot reach the storage service, which we assume is highly unlikely since the storage service is designed to be highly available. 

\subsection{Failure and Recovery} \label{ssec:failure}

This section discusses the behavior of \name when failures occur. For simplicity, we assume each site has one process and we discuss cases where only one site fails at a time. 
\tabref{tab:coord-failure} and \tabref{tab:part-failure} discuss the system behavior of a coordinator and participant, respectively --- when each one fails and when it recovers. The tables also describe the behavior of the failed node after it is recovered. 

\begin{table*}[t]
\begin{tabularx}{\textwidth}{p{1.8in}p{3.9in}X}
\hline
\textbf{Time of Coordinator Failure} & \textbf{Effect of Failure} & \textbf{During Recovery} \\\hline
Before \name starts & Participants (if any) time out and unilaterally abort the transaction. & No action needed \\
\hline
After sending some vote requests 
& Participants that did not receive the request time out and abort unilaterally. Participants that receive the request time out waiting for the decision and execute the termination protocol, which aborts the transaction. & No action needed \\
\hline
After sending all vote requests but before sending any decision 
& All participants time out while waiting for the decision and execute the termination protocol to learn the outcome. & No action needed \\ 
\hline 
After sending some decisions & Participants that did not receive the decision run termination protocol to learn the decision. & No action needed \\
\hline
After sending all decisions & No Effect & No action needed \\
\hline
\end{tabularx}
\caption{Effects of Coordinator Failures.}
\vspace{-.1in}
\label{tab:coord-failure}
\end{table*}
   
\vspace{.05in}
\noindent\textbf{Coordinator Failure}
\vspace{.05in}

Here we list the system behaviors when the coordinator fails at different points of the \name protocol. 

\textbf{Case 1:} The coordinator fails before the protocol starts. In this case, a participant times out waiting for \texttt{VOTE-REQ} (line 13 in \algoref{alg:protocol}). Therefore, all participants can unilaterally abort the transaction locally. 

\textbf{Case 2:} The coordinator fails after sending some but not all vote requests. A participant that did not receive the request has the same behavior as Case 1; namely, it unilaterally aborts the transaction. 
A participant that received the request logs the vote, sends a response to the coordinator, and times out while waiting for the final decision because the coordinator failed (line 21 in \algoref{alg:protocol}). Then the participant runs the termination protocol to check the votes of all participants. 
It either learns the abort decision or appends an \texttt{ABORT} to their logs, thereby aborting the transaction.

\textbf{Case 3:} The coordinator fails after sending all the vote requests but before sending any decision. 
In this case, all participants will time out waiting for the decision from the coordinator. They all run the termination protocol to learn the final outcome of the transaction and act accordingly. 

\figref{fig:1pc_block} illustrates such a case. The figure can be compared with \figref{fig:2pc_block}
showing how \name avoids blocking. After a participant's timeout, instead of contacting the coordinator, which has failed, it contacts all the logs in the shared storage using the \logfunc function. Since all have \texttt{VOTE-YES} in their logs, each participant learns the decision of \texttt{COMMIT} and avoids blocking. 

\textbf{Case 4:} The coordinator fails after sending out the decision to some but not all participants. For participants that have already received the decision, their local  protocol  terminates. The other participants time out waiting for the decision and run the termination protocol to learn the final decision. 

\textbf{Case 5:} The coordinator fails after sending out the decision to all participants. In this case, all participants have completed the local protocol and thus have no effect. 

\newedit{Regardless of the cases, the coordinator has no actions during the recovery since it keeps no state and participants can terminate the transaction on their own.}
    
\begin{table*}[t]
\begin{tabularx}{\textwidth}{p{1.6in}p{2.5in}X}
\hline
\textbf{Time of Participant Failure} & \textbf{Effect of Failure} & \textbf{During Recovery} \\\hline
Before receiving the vote request & The coordinator will timeout, running the termination protocol to abort the transaction & Abort the transaction \\
\hline
After receiving the vote request, before logging vote & Same as above & Same as above\\ 
\hline
After logging vote, before replying to coordinator & The coordinator will run the termination protocol to see the vote and learn the final outcome. & Abort the transaction if local vote is abort; otherwise run termination protocol to learn the outcome. \\
\hline
After replying vote to the coordinator 
& No effect & If decision log exists, follow the decision. Otherwise, same as above. \\
\hline
\end{tabularx}
\caption{Effect of a Participant Failures.}
\label{tab:part-failure}
\vspace{-.3in}
\end{table*}

\vspace{.05in}
\noindent\textbf{Participant Failure}
\vspace{.05in}

We discuss the effects of a participant failing at different points of the \name protocol. 

\textbf{Case 1:} The participant fails before receiving the vote request from the coordinator. In this case, the coordinator times out waiting for all responses from participants and then runs the termination protocol (line 7 in Algorithm 2).

The coordinator logs an \texttt{ABORT} record for the failed participant, thereby aborting the transaction. The coordinator then broadcasts the decision to the remaining participants. If another participant also times out and initiates the termination protocol, the end effect is the same. 
After the failed participant recovers, it runs the termination protocol to learn the final decision. 

\textbf{Case 2:} The participant fails after it receives the vote request but before logging its vote. The behavior of the system is the same as in Case 1 because, from the other participants' perspectives, the behavior of the failed participant is identical. 

\figref{fig:1pc_abort} shows an example of this case. The coordinator receives a \texttt{VOTE-YES} from participant 2 and times out while waiting for a response from participant 1. 
At this point, the coordinator runs the termination protocol to try to log \texttt{ABORT} on behalf of the participants via \logfunc. 
After the coordinator logs \texttt{ABORT} for participant 1 and learns that participant 2 logged \texttt{VOTE-YES}, it logs its abort decision and sends it to participant 2. The example assumes the coordinator times out. Participant 2 might also experience a timeout, which will lead to the same final outcome. 

\textbf{Case 3:} The participant fails after it logs the vote, but before replying to the coordinator. In this case, the coordinator times out waiting for votes and runs the termination protocol. Then it can see all the participants' votes from the storage and learns the outcome. The remaining participants learn the decision either from the coordinator or by running the termination protocol themselves. After the failed participant recovers, it aborts the transaction if it found its local vote is an abort; otherwise, it runs the termination protocol to learn the outcome.

\textbf{Case 4:} The participant fails after sending out the vote. This failure does not affect the others --- the coordinator and remaining participants execute the rest of the protocol normally. After the failed participant recovers, it either finds that it already logged a decision or else runs the termination protocol to learn the outcome.

\subsection{Proof of Correctness} \label{ssec:proof}

This section formally proves the correctness of \name. Our proof follows the structure used in~\cite{bernstein1987concurrency} where an atomic commit protocol is proved as five separate properties 
(AC1--5, see below). We start by defining a \textit{global decision} of a distributed transaction. 

\vspace{.05in}
\noindent \textbf{Definition 1} [Global Decision]: A transaction reaches 
a \texttt{COMMIT} global decision if all participants have logged \texttt{VOTE-YES}; it reaches an \texttt{ABORT} global decision if any participant has logged \texttt{ABORT}. Otherwise the decision is undetermined.
\vspace{.05in}

We first introduce and prove the following lemma.

\vspace{.05in}
\noindent\textbf{Lemma 1} [Irreversible Global Decision]: Once a global decision is reached for a transaction, the decision will not change. 

\noindent\textbf{Proof:} 
There are two cases to consider. In case 1, an abort global decision has been reached meaning one log contains an \texttt{ABORT} record. According to the semantics of \logfunc, no \texttt{VOTE-YES} can be appended to that log anymore, meaning that the global decision cannot switch to commit. In case 2, a commit global decision is reached and all participants have \texttt{VOTE-YES} in their logs. The only way to append an \texttt{ABORT} record is through the termination protocol. But the semantics of \logfunc implies it will not append \texttt{ABORT} because \texttt{VOTE-YES} already exists.\qed
\vspace{.05in}


We now prove the five properties in order.

\vspace{.05in}
\noindent\textbf{Theorem 1} [AC1]: The decision of each participant is identical to the global decision.

\noindent\textbf{Proof:}
\newedit{A participant reaches a decision through either learning the decision from the coordinator or directly seeing or enforcing the vote in each log through the termination protocol. In both cases, a local commit decision is reached only when all logs contain \texttt{VOTE-YES} and an abort local decision is reached only when at least one log contains \texttt{ABORT} (which may be written through the termination protocol), so that the local decision is identical to the global decision.}

\vspace{.05in}
\noindent\textbf{Theorem 2} [AC2]: A participant cannot reverse its decision after it has reached one.\\
\noindent\textbf{Proof:} Due to Lemma 1, once a global decision is reached, it cannot reverse. Due to Theorem 1, each participant will reach the same decision as the global decision, finishing the proof.
\qed
\vspace{.05in}

The following two properties must hold for correctness of 2PC~\cite{bernstein1987concurrency}.

\noindent \textbf{AC3:} The \textit{commit} decision can only be reached if all participants voted Yes. 

\noindent \textbf{AC4:} If there are no failures and all participants voted Yes, then the decision will be to \textit{commit}. 

For the proof of \name, we combine the two properties into the following theorem. 

\vspace{.05in}
\noindent\textbf{Theorem 3} [AC3\&4]: The decision of a transaction is a \textit{commit} if and only if all participants vote Yes and write \texttt{VOTE-YES} to their corresponding logs. \\
\textbf{Proof:} By Definition 1, a transaction’s \textit{global decision} is a \textit{commit} if and only if all participants write \texttt{VOTE-YES} to their logs. By Theorem 1, the decision of each participant is identical to the global decision, finishing the proof. \qed
\vspace{.05in}

Finally, 2PC requires the following property. 

\noindent \textbf{AC5:} Consider any execution containing only failures that the algorithm is designed to tolerate. At any point in this execution, if all existing failures are repaired and no new failures occur for sufficiently long, then all processes will eventually reach a decision. 

For \name, we prove the following theorem which 
achieves a stronger property. 

\vspace{.05in}
\noindent \textbf{Theorem 4} [AC5]: If the storage layer is fault tolerant, then with any failures in the compute layer, the remaining participants will reach a decision in bounded time without requiring the failed sites to be recovered. 

\noindent\textbf{Proof:} Since the storage layer is fault tolerant, once a global decision is reached, an active participant can always learn the global decision through the termination protocol. The only case where a decision cannot be reached is when one participant fails to log its vote. In this case, a coordinator or participant that experiences a timeout will run the termination protocol and directly write an \texttt{ABORT} into the pending participant’s log, which enforces a global decision. 
\qed
\subsection{Read-Only Transactions} \label{ssec:opt}
Conventional 2PC can be optimized for read-only participants~\cite{mohan1986transaction}.
Specifically, if a transaction issues only read requests to one participant, this participant does not need to log during the prepare phase and can directly release locks. For simplicity, we will call such participant a read-only participant. In \name, this optimization has a small subtlety that we will explain in the following two cases.

The simple case is that the entire transaction is read-only and the coordinator knows this fact before starting 2PC. In this case, similar to 2PC, all participants can skip logging in the prepare phase in \name. In practice, we believe it is possible to learn read-only transactions before starting 2PC in many scenarios. 

In the second case, the transaction does not know whether each participant is read-only before starting 2PC. In this case, all partitions including the read-only ones must log \prepare in \name (in contrast to 2PC which skips logging for read-only partitions).
Such a log is necessary because otherwise other participants will interpret the absence of a \prepare in a read-only participant's log as an abort (e.g., consider a read-only transaction that times out waiting for a \texttt{VOTE-REQ} and aborts). Although writing such a log for read-only participants is additional overhead, it can happen in parallel with log writes of read-write participants. 
Therefore, it does not increase the number of log writes on the critical path, though it might affect tail latency if a read-only participant is slow.
By contrast, in 2PC, the coordinator has an extra log write on the critical path, to log the commit decision.
Therefore, \name still has lower latency compared to conventional 2PC.

\subsection{Further Optimization Opportunities}
\label{ssec:further-opt}

With \name, we leverage the functionality that is \emph{already supported in existing storage systems} to optimize 2PC, namely, \logfunc through compare-and-swap. In this subsection, we discuss other optimization opportunities (e.g., some designs in Paxos Commit discussed in \secref{ssec:storage_disag}) if storage systems supports further functionalities.

\textbf{Optimization \#1.} Upon receiving a request, an existing storage service would send the response only to the requesting participant. By letting the storage service respond to multiple participants, we can further reduce the latency of the protocol. 
Specifically, after a participant's vote is logged in the storage layer, the response can be sent to both the requesting participant and the coordinator.
As a result, the coordinator can learn the votes directly from storage instead of requiring another message hop from the participant. 
This saves one message delay in the critical path, without introducing any extra messages.

\textbf{Optimization \#2.} The optimization above can be extended by having the storage broadcast its vote to all participants of a transaction. 
Then each participant can learn the final decision directly from distributed votes, rather than waiting for the coordinator to send a decision message after it receives all votes.
This further reduces the overall latency but does not impact the user-observed latency. Unlike optimization \#1 above, this optimization incurs extra network messages due to broadcasting.

These two optimizations can be implemented as extensions to existing storage services, without changing the underlying replication or consensus protocol.
If the internals of the storage layer are exposed to the compute layer, further optimizations can be enabled, as in 
Paxos Commit as shown in Section 5.6. 
However, this exposure is contrary to our goal of having 
2PC optimizations work with any storage service as long as they support the needed APIs.

\section{Deployment} \label{sec:deployment}

In this section, we discuss the deployment of \name in practical cloud storage systems. In particular, this requires implementing \textit{Log()} and \logfunc introduced in Section~\ref{ssec:storage_api}, using existing cloud storage services. This entails two requirements.

One requirement is to guarantee that after the decision is made it will not be altered, even if multiple participants concurrently execute the termination protocol. We can leverage existing compare-and-swap-like APIs in cloud storage services to support this feature. The other requirement is access control. In 2PC, logging in the prepare phase persists both transaction states and the user data. 
In \name, a participant may need to access other participants' transaction states. This requires separate access control for transactions' states and user data so that a participant cannot read others' log of user data while accessing the transaction states. 

In this paper, we studied a wide range of modern in-cloud storage services 
on how these services support compare and swap (CAS) and data privacy, and we deploy \name on two of them --- Redis and Azure Blob Storage (a.k.a. Windows Azure Storage for Blob) for quantitative evaluations in next section. 

\subsection{Deployment on Redis}
Redis~\cite{redis} is an in-memory data store that supports optional durability and can be used as a distributed, in-memory key-value database, cache, and message broker. 

\subsubsec{Compare-and-swap}
Redis supports compare-and-swap operations through the “EVAL” command and Lua scripting~\cite{redis-lua}. It guarantees that Lua scripts run by “EVAL” commands are executed atomically. That is, the effect of a script is either all or none with respect to scripts and commands of other Redis clients. 
\lstref{code:redis-log-if-ne} shows an implementation of \logfunc using cpp\_redis~\cite{redis-cpp}, an asynchronous multi-platform lightweight Redis client for C++. 

\begin{lstlisting}[showstringspaces=falselanguage=c, basicstyle=\scriptsize\ttfamily,
        escapeinside={@}{@}, numbers=left, caption={Implementation of Synchronous \logfunc on Redis}, label={code:redis-log-if-ne}]
auto lua_script = R"(
	redis.call('set', KEYS[1], ARGV[1])
	redis.call('setnx', KEYS[2], ARGV[2])
	local state = tonumber(redis.call('get', KEYS[2]))
	return {tonumber(state)};
)";
string id = to_string(participant_id) + "-" + to_string(txn_id);
vector<string> keys = {"data-" + id, "state" + id};
vector<string> args = {data, to_string(VOTE_YES)};
client.eval(lua_script, keys, args, [](reply& reply) {
	// a callback function to execute when reply is ready
	uint64_t txn_state = response.as_array()[0].as_integer();
});
client.sync_commit();
\end{lstlisting}


\subsubsec{Access Control}
Redis Access Control List (ACL)~\cite{redis-acl} manages users’ access to certain commands and/or keys patterns. 
We can configure that each participant has read-write access to all participants' transaction states but cannot access other participants' user data. 
We do this using the “ACL SETUSER” command to set up read-write access to all keys starting with the pattern “data-<current participant id>” and “state-” if using the implementation in~\listref{code:redis-log-if-ne}.

\subsection{Deployment on Microsoft Azure Blob Storage}

Microsoft Azure Blob Storage is a scalable storage system that supports secure object storage for cloud-native workloads, archives, data lakes, high-performance computing, and machine learning.

\subsubsec{Compare-and-swap}
Azure Storage assigns an identifier to each stored object. The identifier is updated every time the object is updated. An HTTP GET request returns the identifier as part of the response using the Etag (entity tag) header defined within HTTP. A user updating an object can send in the original Etag with an “If-Match” conditional header so that the update will be performed only if the stored ETag matches the one passed in the request~\cite{azure-etag}. 

\begin{lstlisting}[showstringspaces=falselanguage=c, basicstyle=\scriptsize\ttfamily,
        escapeinside={@}{@}, numbers=left, caption={Implementation of Synchronous \logfunc on Azure Blob with Separate ACLs}, label={code:azure-log-if-ne}]
string id = to_string(participant_id) + "-" + to_string(txn_id);
if (data) {
    // request issued during normal execution (non-failure)
    storage::cloud_block_blob data_blob =
        container.get_block_blob_reference(U("data-" + id));
    auto t = data_blob.upload_text(U(data));
}
storage::cloud_block_blob state_blob =
    container.get_block_blob_reference(U("state-" + id));
try {
    storage::access_condition condition =
        storage::access_condition::generate_if_not_exists_condition();
    storage::blob_request_options options;
    storage::operation_context context;
    state_blob.upload_text(U(state), condition, options, context);
} catch (const std::exception &e) {
    // update failed, log already exist
    State state = (State) stoi(blob_status.download_text());
}
\end{lstlisting}

\subsubsec{Access Control}
Azure Blob Storage supports Azure attribute-based access control (Azure ABAC). It allows read access to blobs based on tags and custom security attributes. As there is no transaction support, we use two separate requests to log the transaction state after successfully logging the actual data as shown in \lstref{code:azure-log-if-ne}. 
In this case, \name shows no improvements over 2PC as we will show in \figref{fig:blob_num_nodes_e}. Many applications do not require the data to be private to the corresponding partitions, in which case \name does not introduce this extra overhead. 

\subsection{Deployment on Key-Value Databases}


Amazon DynamoDB~\cite{decandia2007dynamo} is a highly available key-value storage system with rich APIs~\cite{dynamo-cas}. The “putItem” and “updateItem” APIs provide conditional puts and updates respectively to achieve CAS functionality. The “TransactWriteItems” API allows submitting multiple actions in one request including the “putItem” and “updateItem” mentioned above. As DynamoDB offers item-level access control~\cite{dynamo-acl}, transaction data and transaction states can be kept as separate attributes in a table or kept in different tables, 
while they can be updated simulatenously using a single "TransactWriteItems" request.

Google Cloud Bigtable~\cite{chang2008bigtable} is a Distributed Storage System for Structured Data. It supports conditional writes~\cite{google-cas} so \name can implement \logfunc on top. 
Google Cloud uses Identity Access Management (IAM) for access control. It is a role-based access control mechanism. For Bigtable, it can control accesses at the table level so that transaction states and logs of user data can be kept in separate tables. Each participant can have read/write permission to its own log of user data stored in one table, as well as all transaction states stored in another table.
However, Bigtable does not support writing to multiple tables in a batch, which, like Azure Storage, causes extra overhead when using it for \name. 
\section{Experimental Evaluation}

In this section, we compare the performance of \name with standard 2PC. We introduce the experimental setup in~\secref{ssec:exp_setup}. We evaluate \name in a range of settings to demonstrate its efficacy in reducing latency, primarily in the commit phase. 

\subsection{Experimental Setup}\label{ssec:exp_setup}

\subsubsection{Architecture} 
In this paper, we focus on an architecture comprised of multiple nodes in the compute layer accessing a disaggregated storage service. The database data is partitioned, where each compute node runs a resource manager and has exclusive access to one partition. 
While a compute node executes transactions, it sends data access requests to the corresponding compute node. At commit time, one compute node coordinates the resource managers participated in the transaction to commit. Every compute node can write log records to the storage service.

We implemented \name on Sundial~\cite{yu2018sundial}, 
an open-source distributed DBMS testbed. 
Our implementation is public\footnote{\url{https://github.com/CloudOLTP/Cornus-Public}}. 
Compute nodes communicate using gRPC~\cite{grpc}, which can be either synchronous or asynchronous. 
Each node has a \textit{gRPC} client for sending requests and
a \textit{gRPC} server that manages a pool of server threads to handle requests.

\subsubsection{Compute Node Hardware and Storage Service} \label{sssec:storage_service}
For compute nodes, we use a cluster of up to eight servers running Ubuntu 18.04 on Microsoft Azure. Each server has one Intel(R) Xeon(R) Platinum 8272CL CPUs (8 cores $times$ 2 HT) and 64 GB of DRAM. The servers are connected by a 12.5 Gbps Gigabit Ethernet. 
Based on our measurements, a network round trip is 0.5 ms between two compute nodes.

\name can use any cloud storage service.
We use the following two services in three different configurations. 
\begin{itemize}[leftmargin=*]
	\item \textbf{Microsoft Azure Blob Storage (Azure Blob)}~\cite{azure-blob}: We use a StorageV2 (general purpose v2) Azure storage account to store blobs, 
	with geo-redundant storage (GRS) replication enabled. Data is stored in two regions. In the primary region, data is synchronously three-way replicated in one physical location. Then the data is asynchronously copied to a secondary region hundreds of miles from the primary.
	The average time for a conditional write request is 10.40 ms and for a plain write request is 10.29 ms.
	\item \textbf{Microsoft Azure Cache for Redis (Redis)}~\cite{azure-redis}: This experiment uses the Redis service provided by Microsoft Azure. We created a Premium P4 Redis instance of version 4.0.14. It uses master-slave replication with one master and one slave in the same region. Only the master node accepts reads and writes. It applies changes to the slave node asynchronously. The average time for a conditional write request is 1.96 ms and for a plain write request is 1.84 ms.
\end{itemize} 

\subsubsection{Workloads} \label{sssec:workloads}

We use the Yahoo! Cloud Serving Benchmark (YCSB) \cite{cooper2010benchmarking} for performance evaluation. YCSB is a synthetic benchmark modeled on cloud services. It contains a single table partitioned across servers in a round-robin fashion. Each partition contains 10 GB data with 1 KB tuples. Each transaction accesses 16 tuples as a mixture of reads (50\%) and writes (50\%). 
The queries access tuples following a Zipfian power law distribution controlled by a parameter $\theta$. By default, we use $\theta=0$, which means data access is uniformly distributed. All transactions are executed as stored procedures that contain program logic intermixed with queries.

\subsubsection{Implementation Details and Parameter Setup} \label{ssec:exp_param}
Unless otherwise specified, we use the following parameter settings: We evaluate the system on up to eight compute nodes and a storage service. Eight worker threads per node execute the transaction logic, and eight worker threads per node serving the remote requests. The default concurrency control algorithm is NO-WAIT. 

For each data point, we run five trials with 30 seconds per trial. We collect the latencies for distributed transactions --- transactions involving more than one partition (node) --- and take the result from the trial with median average latency. Running the experiments for a longer time does not change the conclusions. 

We assume the coordinator of a transaction can learn whether it is read-only at the end of the execution phase. Therefore, \name and 2PC can skip both the prepare and commit phases for read-only transactions~\cite{mohan1986transaction}. 

Details of implementing \logfunc on different storage services are described in Section 4. As Azure Blob does not support batch update of two resources with separate access control, we implemented two versions. In the default version, we used the same access control for transaction data and transaction states for all experiments in this section. The second version with Azure Blob uses separate ACLs. Profiling shows that the second version increases the time for \logfunc from an average of 10.40 ms to an average of 18.43 ms.

\begin{figure}[t]
	\begin{subfigure}[b]{0.25\textwidth}%
    	\center
        \includegraphics[width=\linewidth]{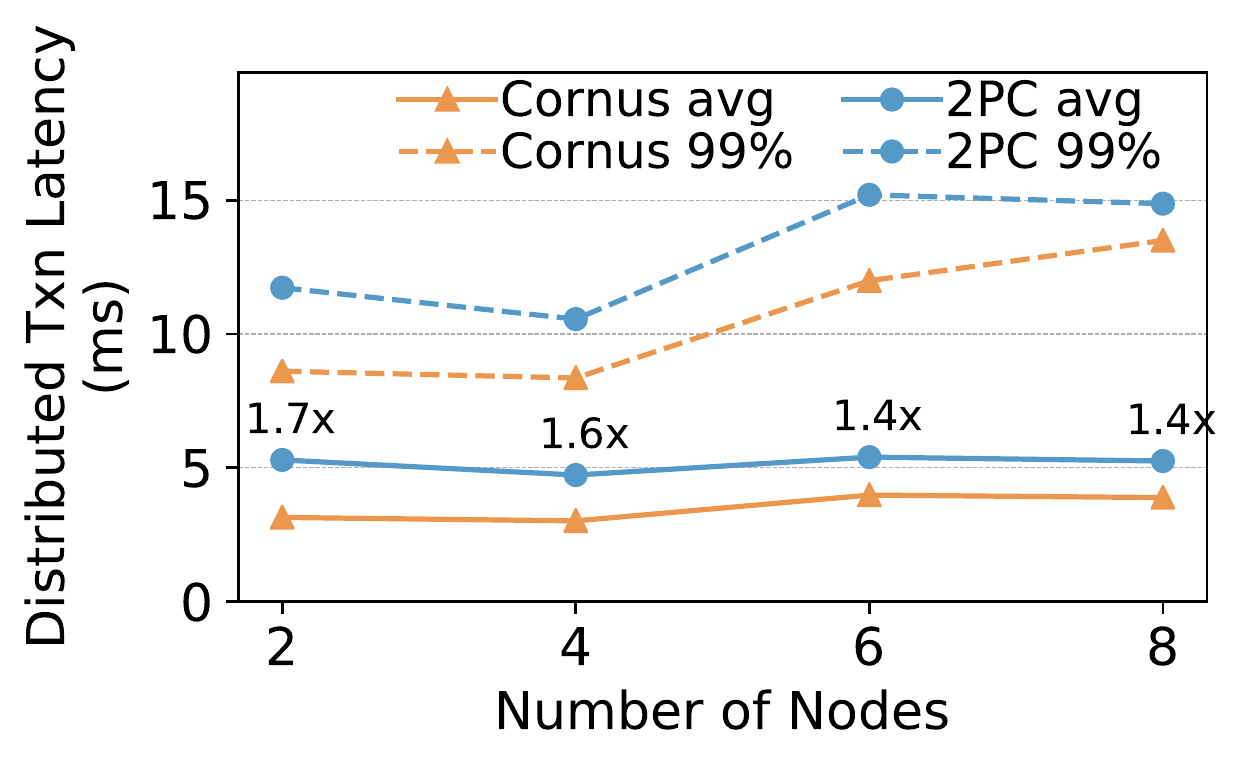}
        \caption{Latency (Redis)}\label{fig:redis_num_nodes_a}
    \end{subfigure}%
    \begin{subfigure}[b]{0.25\textwidth}%
    	\center
        \includegraphics[width=\linewidth]{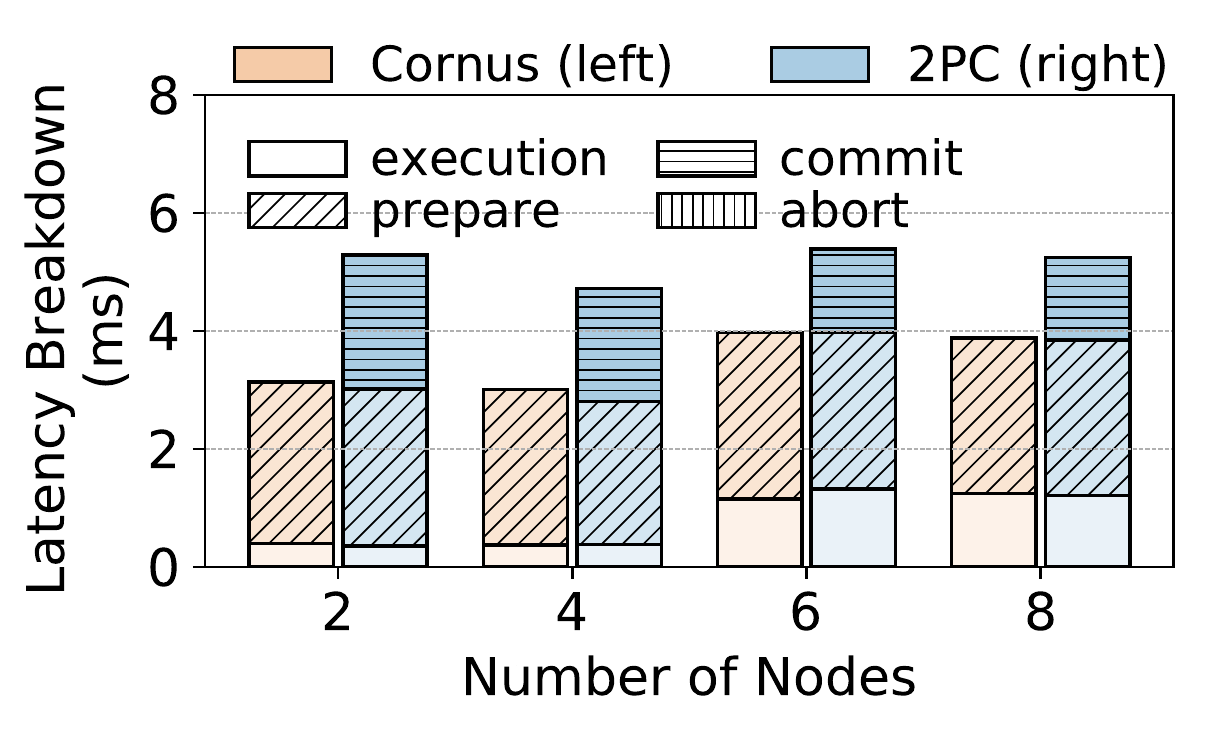}
        \caption{Latency Breakdown (Redis)}\label{fig:redis_num_nodes_b}    
    \end{subfigure} 
    
	\begin{subfigure}[b]{0.25\textwidth}%
    	\centering
        \includegraphics[width=\linewidth]{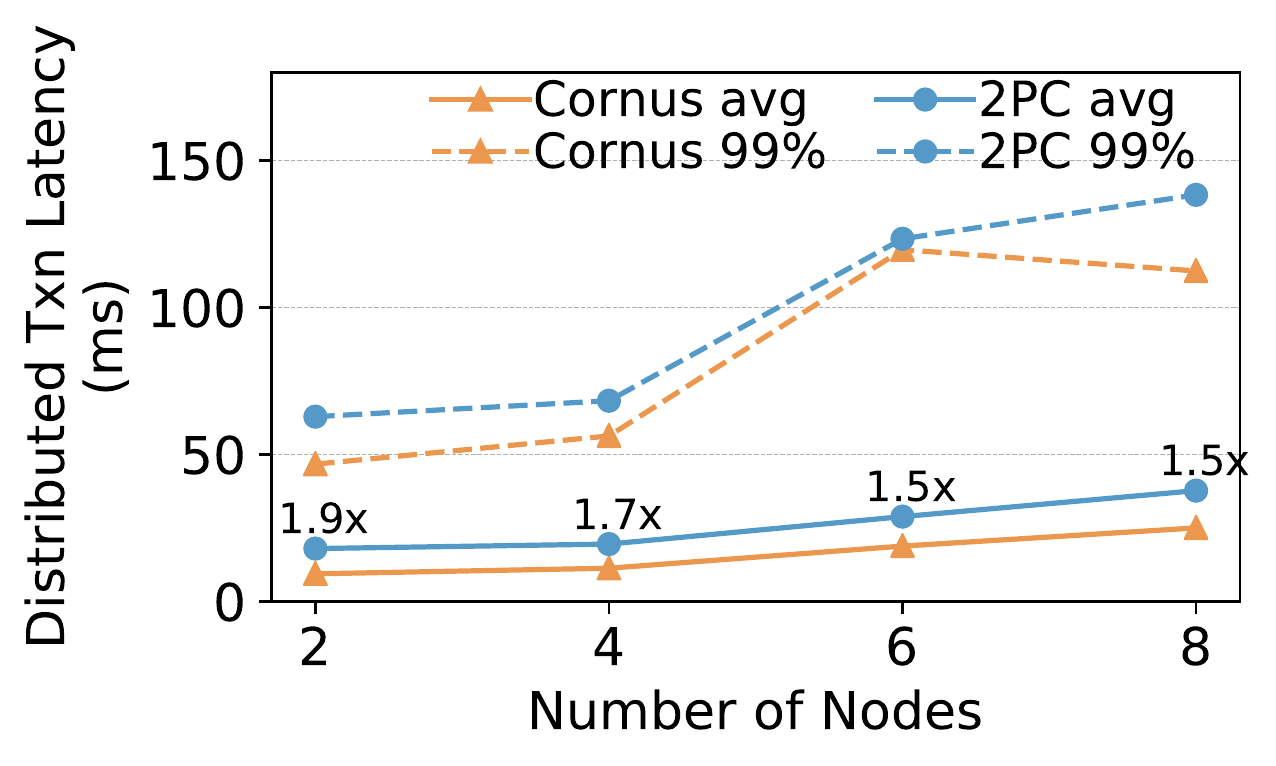}
        \caption{Latency (Azure Blob)}
        \label{fig:blob_num_nodes_c}
    \end{subfigure}%
    \begin{subfigure}[b]{0.25\textwidth}%
    	\centering
        \includegraphics[width=\linewidth]{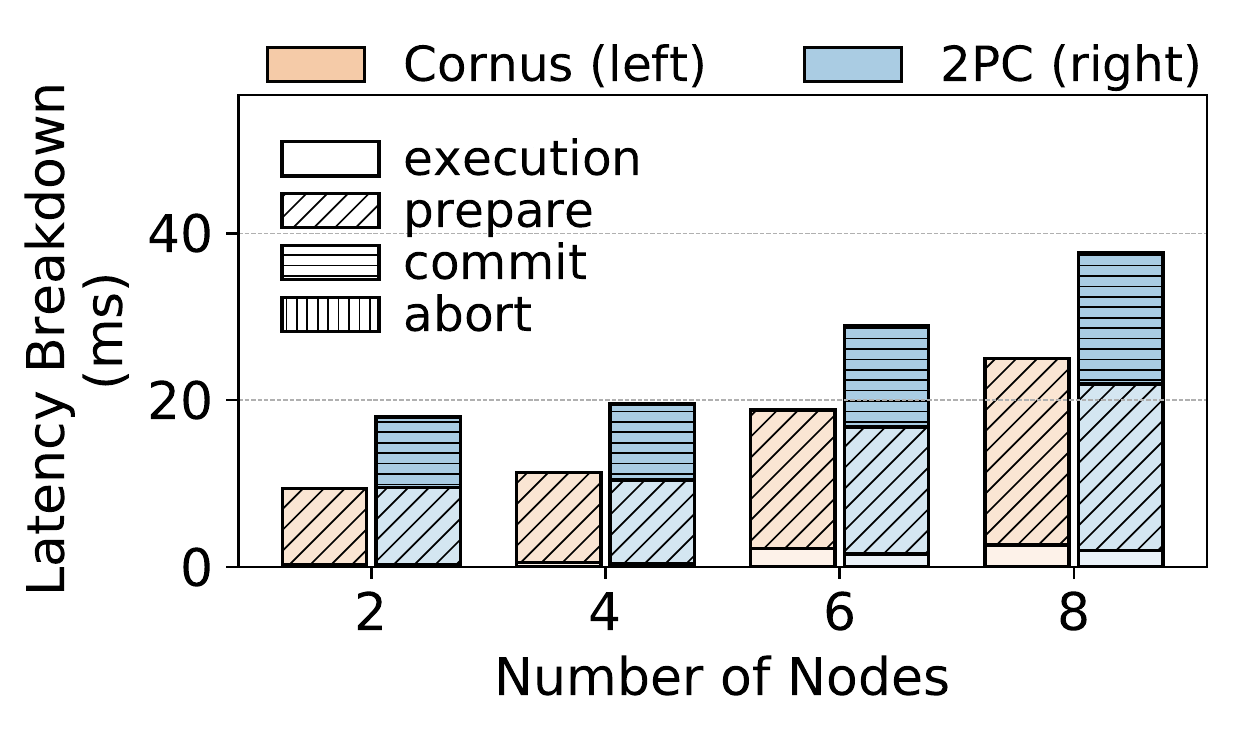}
        \caption{Latency Breakdown (Azure Blob)}\label{fig:blob_num_nodes_d}    
    \end{subfigure}%
    
	\begin{subfigure}[b]{0.25\textwidth}%
    	\centering
        \captionsetup{justification=centering}
        \includegraphics[width=\linewidth]{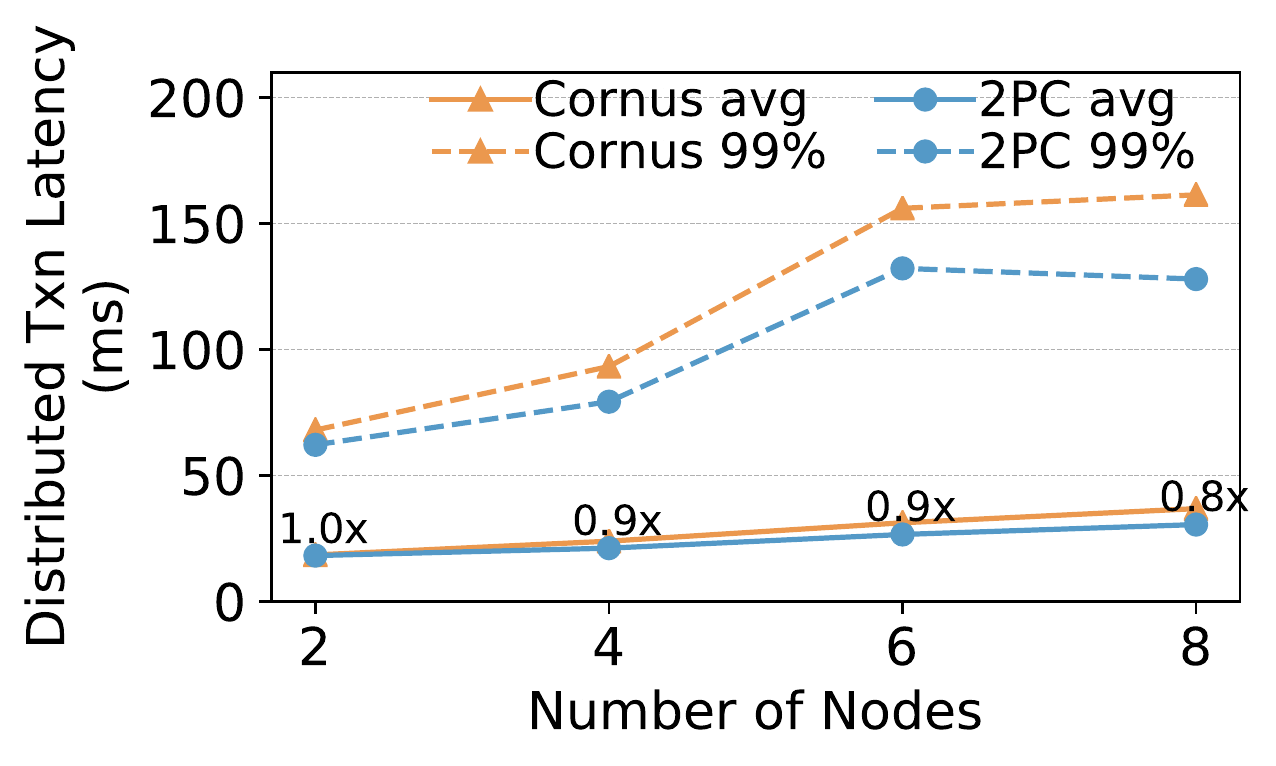}
        \caption[]{ Latency \\
        (Azure Blob - Separate ACLs)}
        \label{fig:blob_num_nodes_e}
    \end{subfigure}%
    \begin{subfigure}[b]{0.25\textwidth}%
    	\centering
        \captionsetup{justification=centering}
        \includegraphics[width=\linewidth]{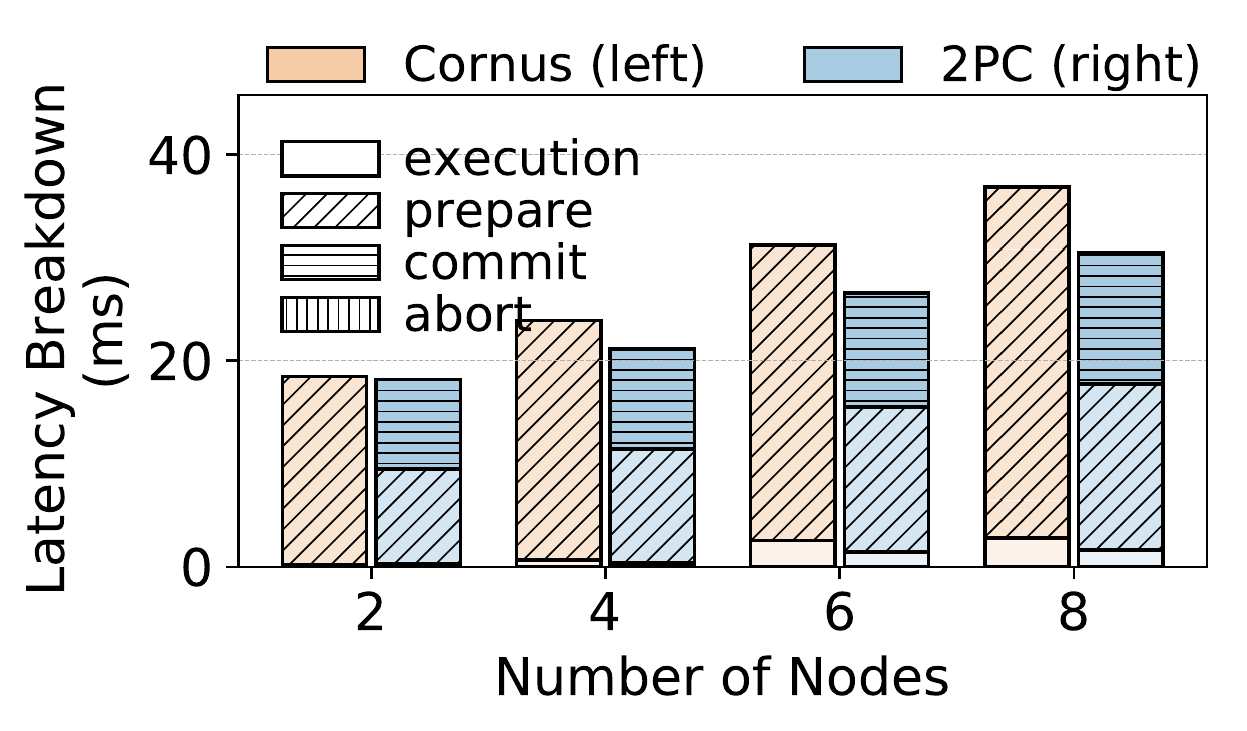}
        \caption{Latency Breakdown\\(Azure Blob - Separate ACLs)}\label{fig:blob_num_nodes_f}
    \end{subfigure}%
    \vspace{-.05in}
    \caption{Scalability}
    \label{fig:scalability}
\end{figure}

\subsection{Scalability}
\indent We first evaluate \name's scalability on YCSB as the number of compute nodes increases from 2 to 8. We set the parameter to default values described in \secref{sssec:workloads}. 
The results in~\figref{fig:scalability}(a--d) show that both \name and 2PC scale well in YCSB with Redis or Azure Blob. As the number of nodes increases, the latency of both 2PC and \name increases linearly. The speedup of \name over 2PC on average latency slightly decreases as the number of nodes increases. This is due to the time spent on RPC calls in execution phase increases. Overall, the latency speedup is up to 1.9$\times$.

\figref{fig:blob_num_nodes_e} and \figref{fig:blob_num_nodes_f} show the evaluation of an implementation on Azure Blob with separate access control for transaction data and transaction states as introduced in \secref{ssec:exp_param}. In 2PC, data and states are stored in a resource within the same access control group as both of them will not be accessed by other partitions. However, in \name, the data and states are stored in separate access control groups so that two remote logging requests instead of one must be used for \logfunc. Thus, \name spent $\sim$9.48 ms more time in the prepare phase for logging than 2PC (\figref{fig:blob_num_nodes_f}) and shows no improvement.
We conclude that the current version of Azure Blob cannot benefit from \name for applications that want separate access control between data and transaction states.

\begin{figure}[t]
    \begin{subfigure}[b]{.25\textwidth}%
        \center
        \includegraphics[width=\linewidth]{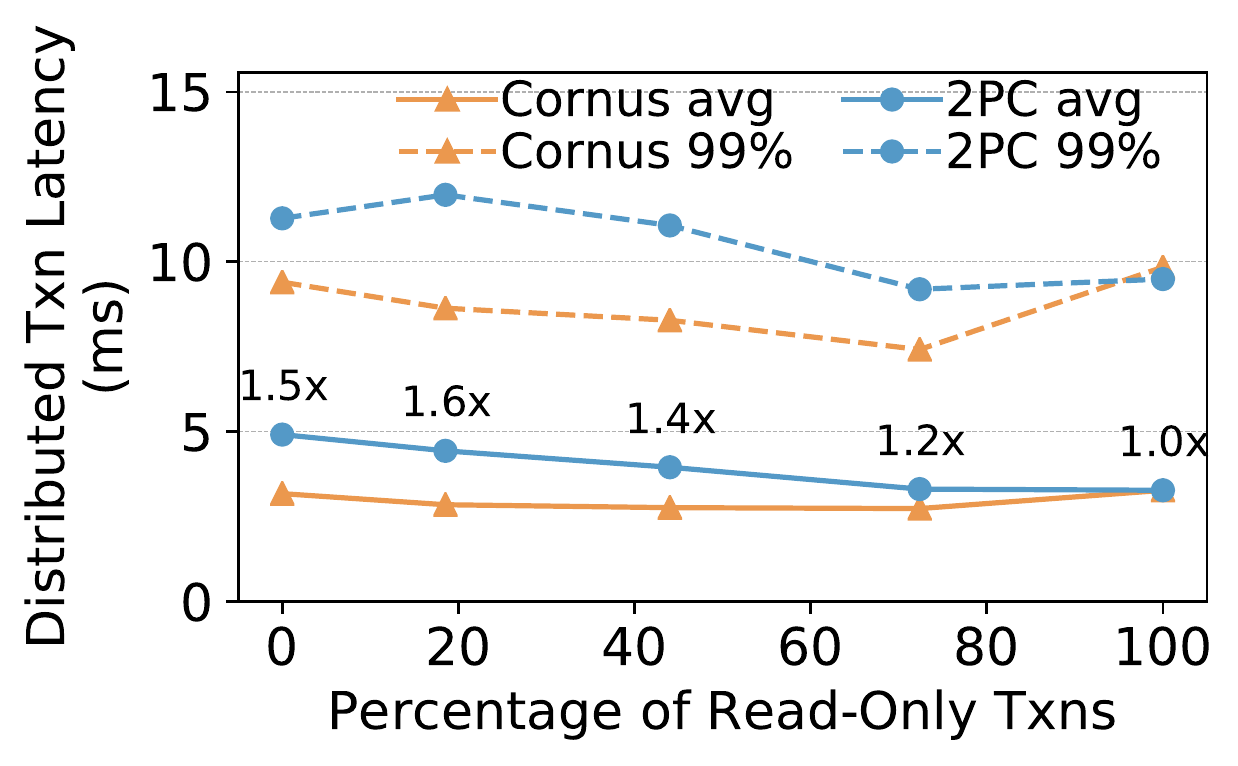}
        \vspace{-.2in}
        \caption{Latency (Redis)}\label{fig:read_ratio_a}
    \end{subfigure}%
    \begin{subfigure}[b]{.25\textwidth}%
        \center
        \includegraphics[width=\linewidth]{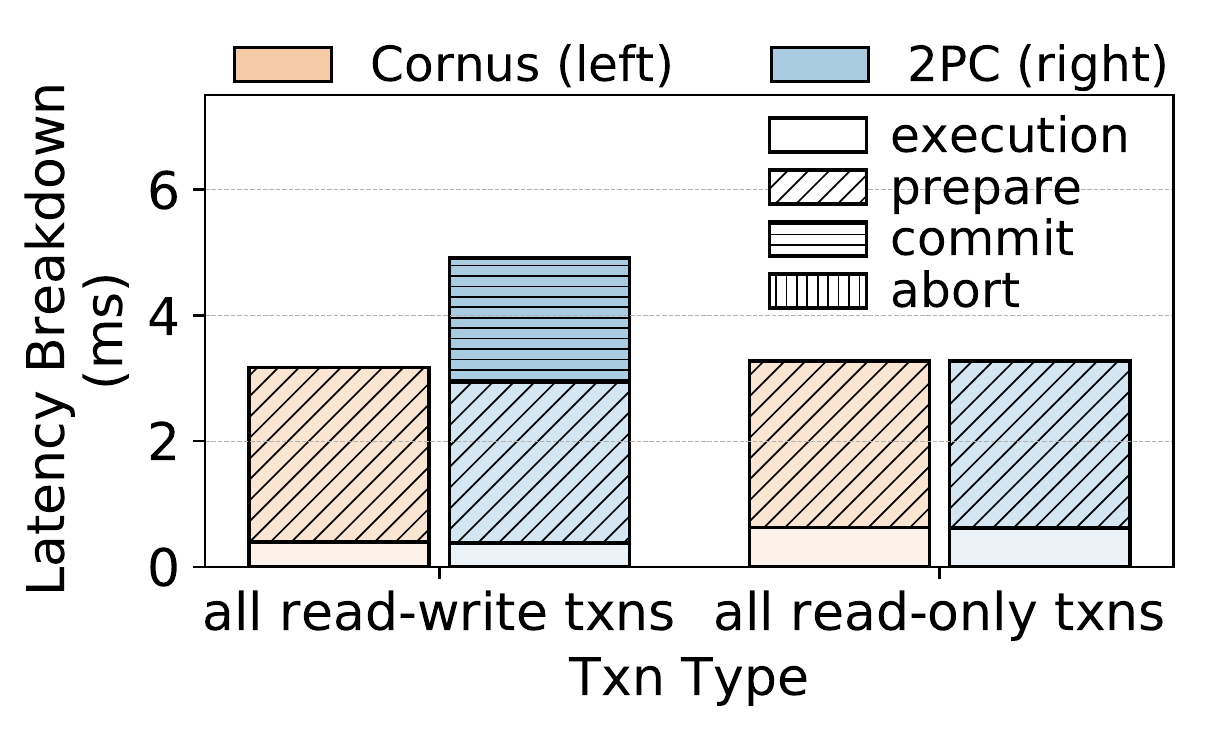}
        \vspace{-.2in}
        \caption{Latency breakdown (Redis)}\label{fig:read_ratio_b}
    \end{subfigure}%
    
    \begin{subfigure}[b]{.25\textwidth}%
        \center
        \includegraphics[width=\linewidth]{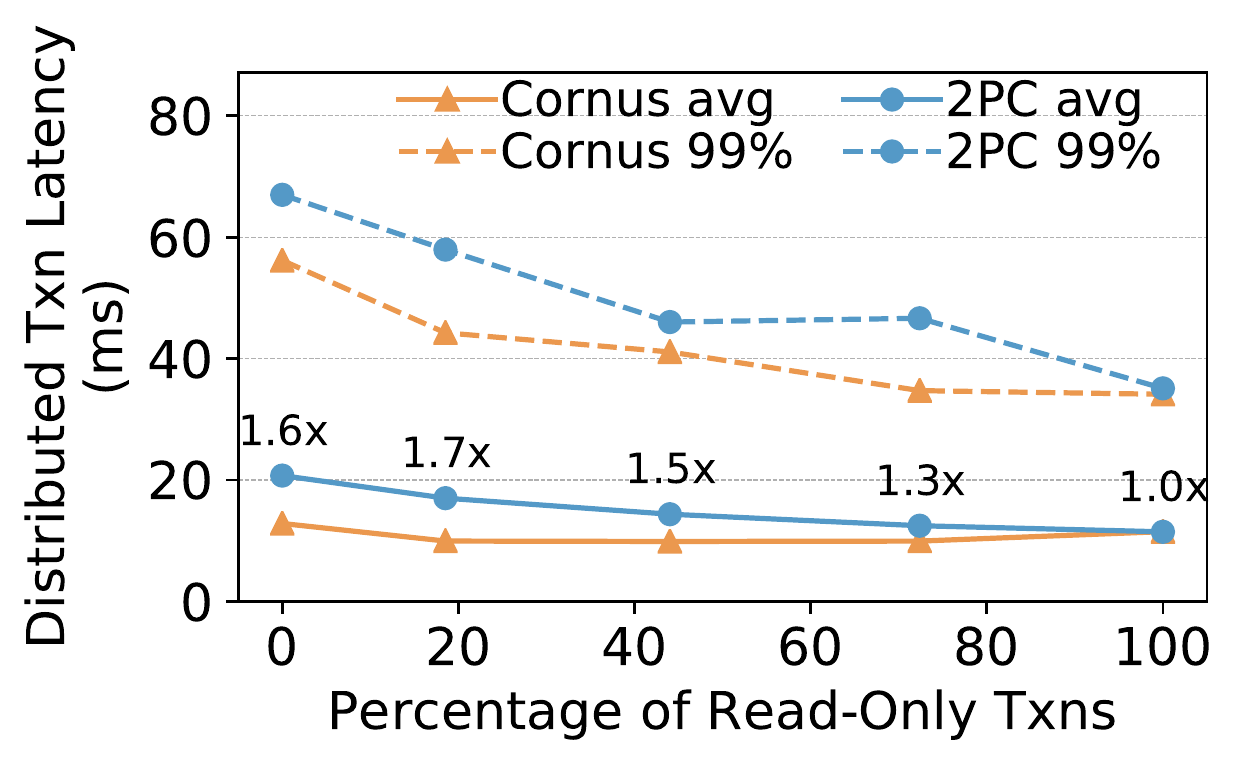}
        \vspace{-.2in}
        \caption{Latency (Azure Blob)}\label{fig:read_ratio_c}
    \end{subfigure}%
    \begin{subfigure}[b]{.25\textwidth}%
        \center
        \includegraphics[width=\linewidth]{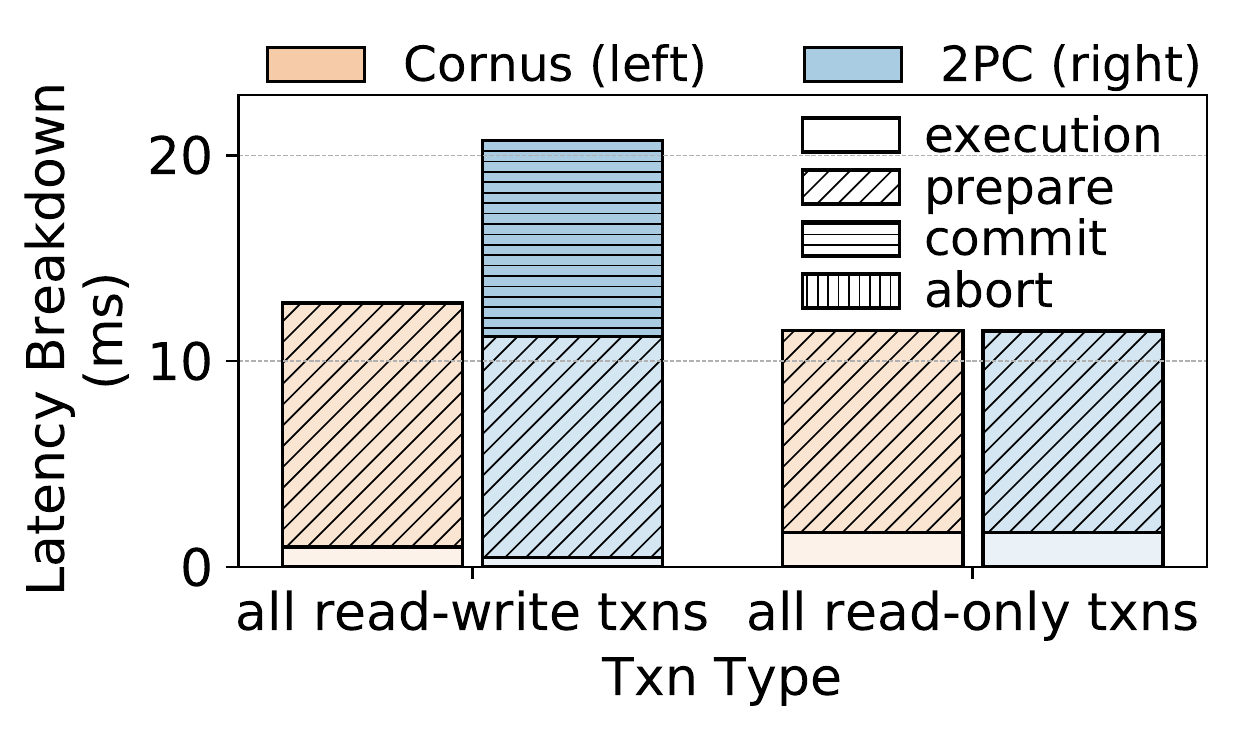}
        \vspace{-.2in}
        \caption{Latency breakdown (Azure Blob)}\label{fig:read_ratio_d}    
    \end{subfigure}%
    \caption{Varying percentage of read-only transactions}

    \label{fig:read_ratio}
\end{figure}


\subsection{Percentage of Read-only Transactions}

We evaluate the performance of \name under YCSB with different fractions of read-only transactions. 
We manage the percentage of read-only transactions by controlling the probability of each request being read in a transaction. With $n$ requests per transaction and each request has a probability of $p$ being read, the percentage of read-only transactions is expected to be $n^p$. We expect \name to obtain a latency speedup solely for read-write transactions 
because both \name and 2PC omit the prepare and commit phases for read-only transactions, as discussed in ~\secref{ssec:opt} and ~\secref{ssec:exp_param}.

The results shown in~\figref{fig:read_ratio_a} 
and ~\figref{fig:read_ratio_c} 
match the expectation. The improvement of \name (relative to 2PC) grows as the percentage of read-only transactions decreases. When there are more read-write transactions, \name improves both average and P99 latency over 2PC by up to 1.7$\times$. The pattern is consistent across different storage services. The result can be explained by the latency breakdown illustrated in ~\figref{fig:read_ratio_b} 
and ~\figref{fig:read_ratio_d}
. \name improves latency for read-write transactions by eliminating the commit phase,  which takes a significant amount of time especially in Azure Blob with geo-distribution and synchronous replication
. Finally, we note that \name spends slightly more time in the prepare phase \newedit{than 2PC due to the subtle differences between \textit{Log()} and \textit{LogOnce()} (Algorithm 1 Line 16)}. 

\begin{figure}[t]
    \begin{subfigure}[b]{.24\textwidth}%
        \center
        \includegraphics[width=\linewidth]{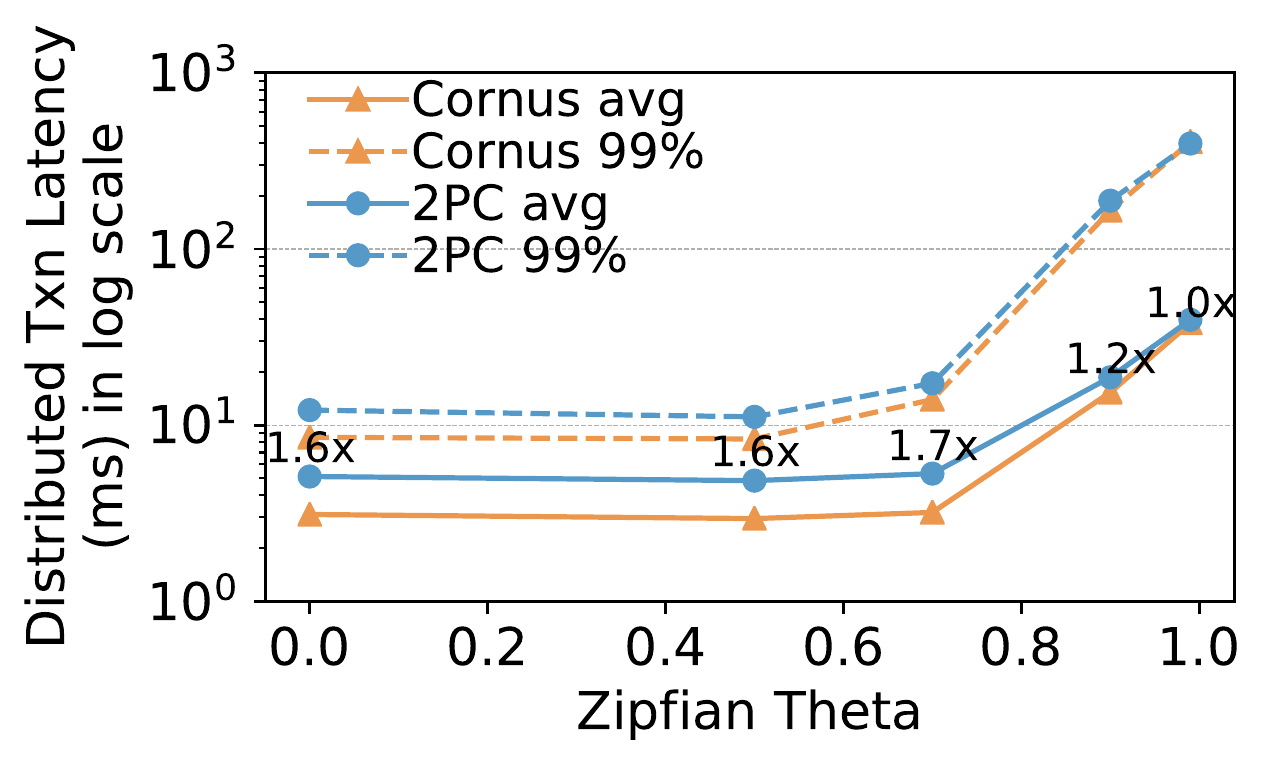}
        \vspace{-.2in}
        \caption{Latency (YCSB, Redis)}\label{fig:redis_zipf_a}
    \end{subfigure}%
    \begin{subfigure}[b]{.26\textwidth}%
        \center
        \includegraphics[width=\linewidth]{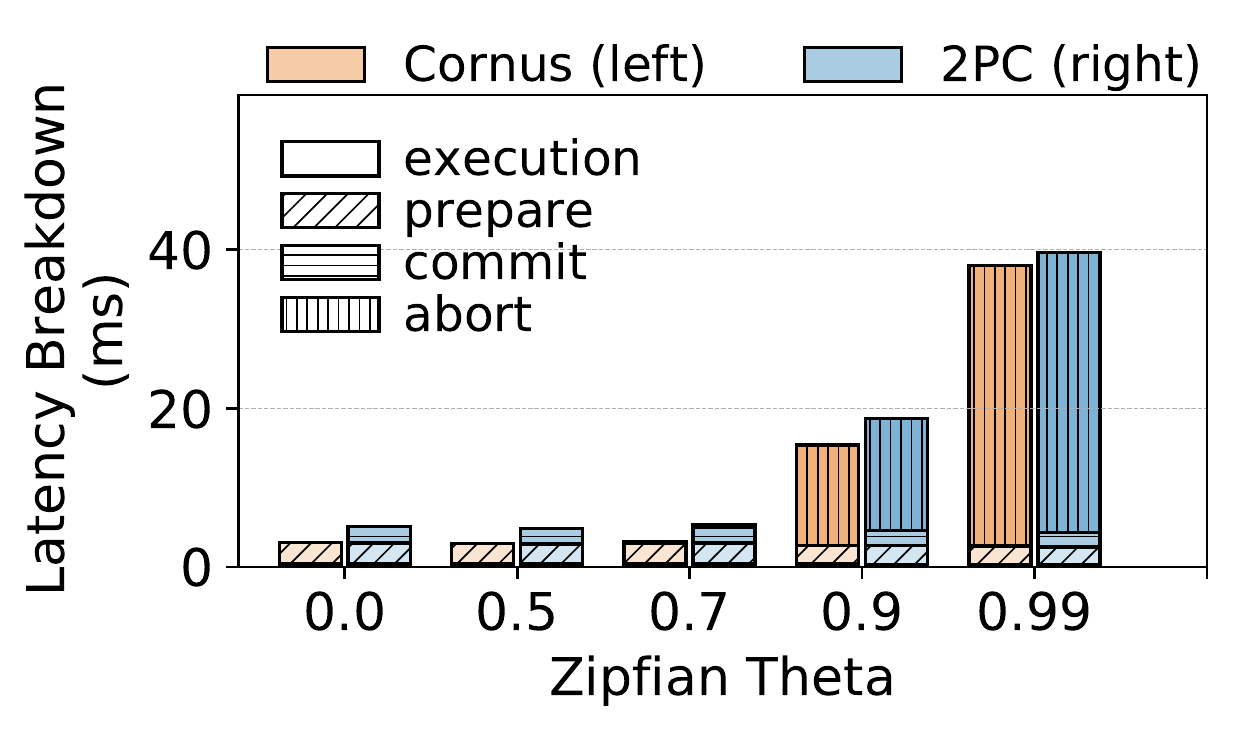}
        \vspace{-.2in}
        \caption{Latency breakdown (YCSB, Redis)}\label{fig:redis_zipf_b}   
    \end{subfigure}%
    
    \begin{subfigure}[b]{.24\textwidth}%
        \center
        \includegraphics[width=\linewidth]{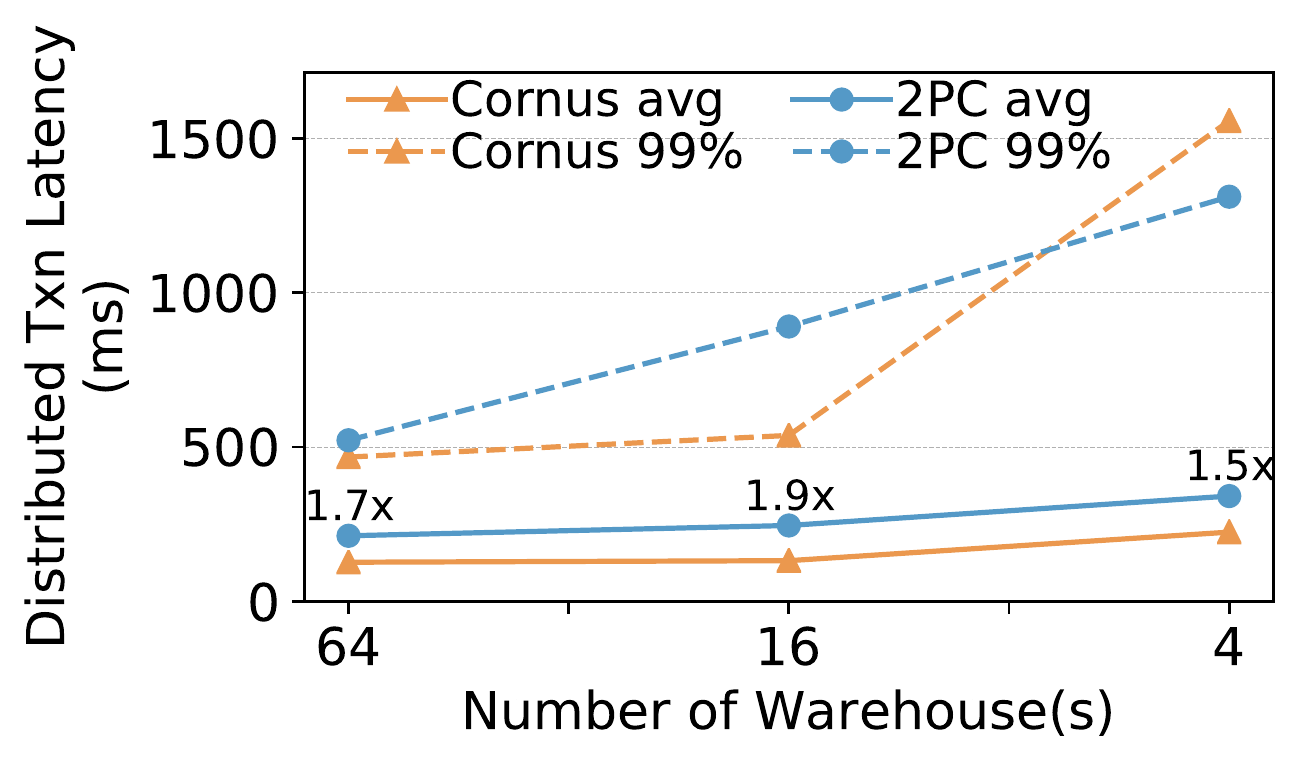}
        \caption{\edit{Latency (TPC-C, Redis)}}\label{fig:redis_tpcc_a}
    \end{subfigure}%
    \begin{subfigure}[b]{.26\textwidth}%
        \center
        \includegraphics[width=\linewidth]{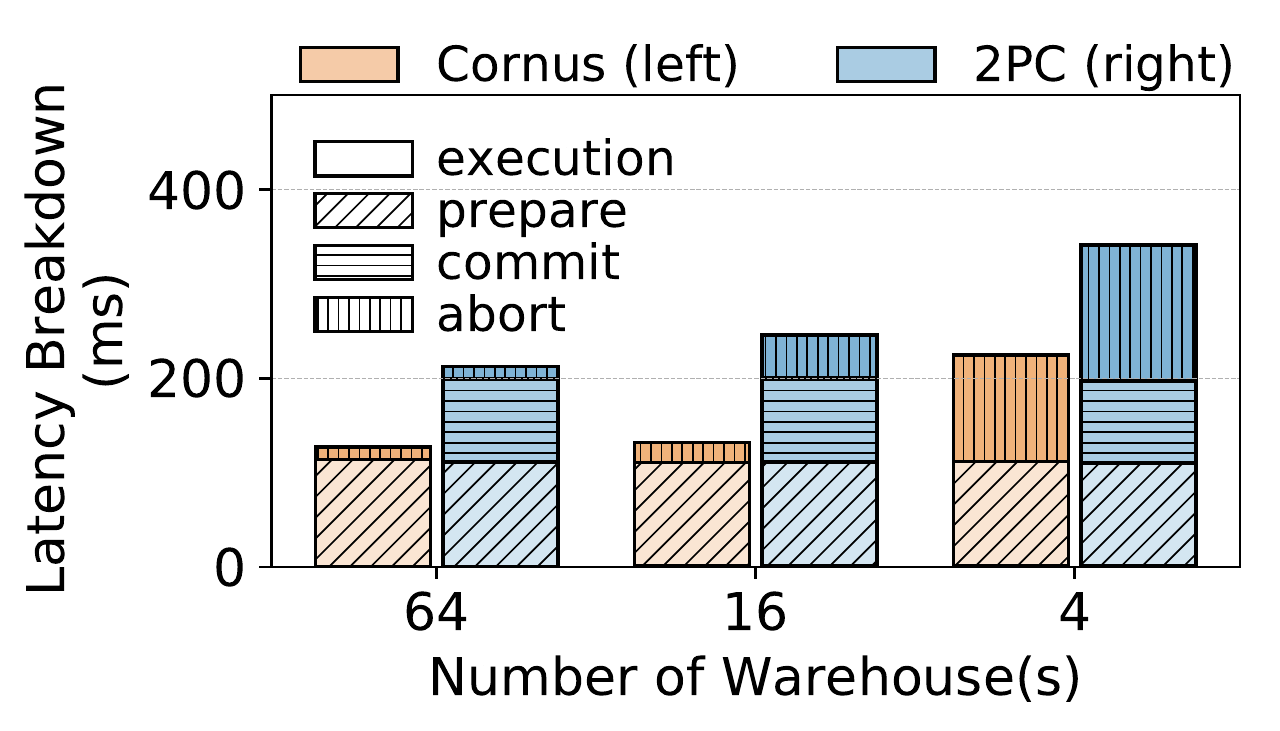}
        \caption{\edit{Latency breakdown (TPC-C, Redis)}}\label{fig:redis_tpcc_b}    
    \end{subfigure}%
    \caption{\edit{Varying workload contention}}
    \vspace{-.1in}
    \label{fig:zipf}
\end{figure}

\subsection{Contention}

We evaluate the performance of \name under varying contention using YCSB and TPC-C workloads.
For YCSB, we adjust the Zipfian distribution of data accesses through $\theta$. Increasing $\theta$ increases the level of contention. 
For TPC-C, we vary the number of warehouses; fewer warehouses indicate higher contention in the workloads.
\figref{fig:zipf} shows the results of both workloads; the x-axis from left to right indicates low to high contention for both YCSB and TPC-C.
As shown in the figure, \name always improves transaction latency over 2PC --- the improvement is up to 1.8 $\times$ for YCSB, and 1.9 $\times$ for TPC-C workloads.
\figref{fig:redis_zipf_b} and \figref{fig:redis_tpcc_b} show that \name provides less improvement under high contention since the abort time dominates the total transaction elapsed time.

\begin{figure}[t]
	\begin{subfigure}[b]{0.48\linewidth}%
    	\center
        \includegraphics[width=\linewidth]{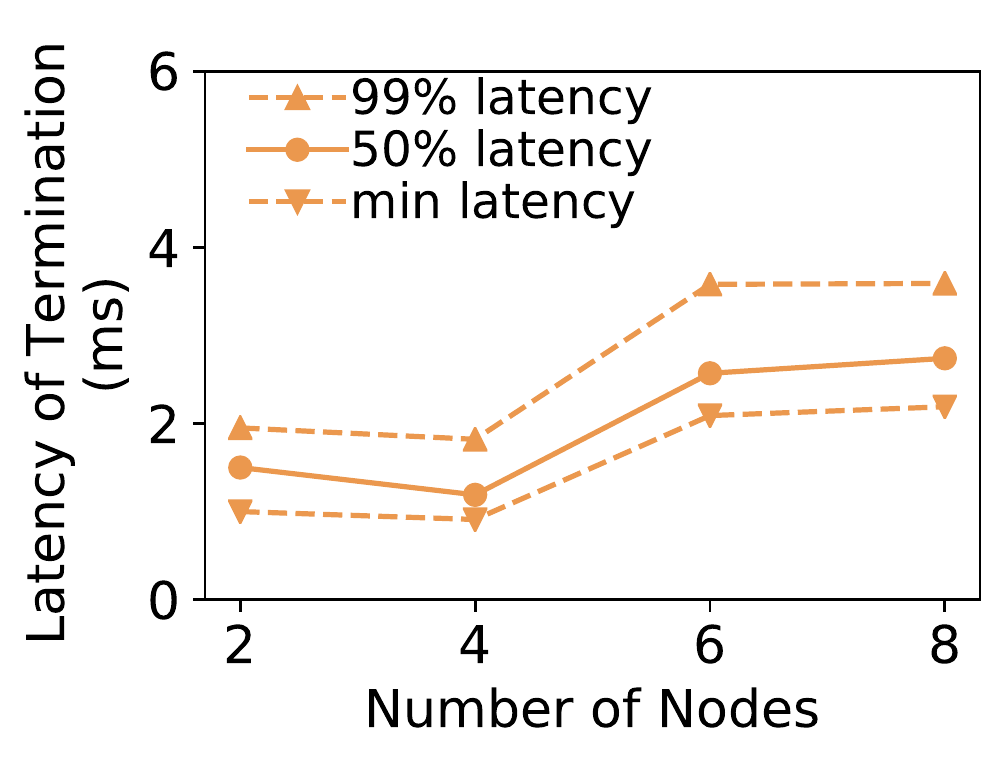}
        \caption{Latency (Redis)}\label{fig:redis_failure_num_nodes_a}
    \end{subfigure}%
	\begin{subfigure}[b]{0.48\linewidth}%
    	\center
        \includegraphics[width=\linewidth]{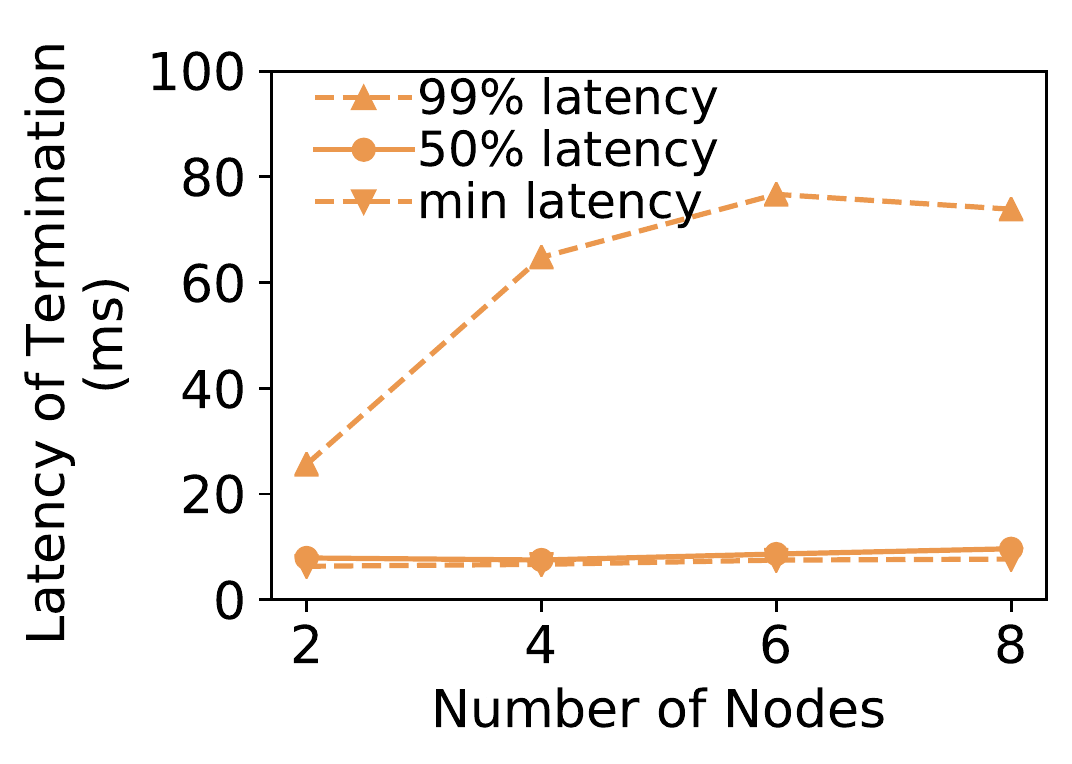}
        \caption{Latency (Azure Blob)}
        \label{fig:blob_failure_num_nodes_a}
    \end{subfigure}%
    \caption{Time to terminate transactions on failure}
    \label{fig:failure}
\end{figure}

\subsection{Time to Terminate Transactions on Failure}
\figref{fig:failure} shows the time to run the termination protocol in \name once it is triggered. In 2PC, there is no bound on the time to run the termination protocol --- a transaction that falls into uncertain states (due to coordinator failure before the decision is sent out to any participants) is blocked indefinitely, until the coordinator recovers. However, in \name, compute node failure does not lead to blocking. In this experiment, we assume the storage is still available and measure the time to terminate a transaction through the termination protocol while varying the number of nodes. As shown, for up to 8 nodes, \name always terminates a transaction within 4 ms with Redis and within 20 ms on average with Azure Blob. The tail latency of Azure Blob increases more than Redis as the number of nodes increases. This is due to the geo-distribution setup and some synchronous replication in Azure, while the two replicas of Redis are co-located in the same region and only perform asynchronous replication, as introduced in \secref{sssec:storage_service}.

\subsection{2PC Optimizations} 
\label{ssec:eval-optimization}

In this section, we evaluate common 2PC optimizations and compare them with Cornus. These optimizations typically make different system/workload assumptions from the classic 2PC; therefore we conduct the comparisons here instead of in the main results. 

\begin{figure}[t]
	\begin{subfigure}[b]{0.48\linewidth}%
    	\center
        \includegraphics[width=\linewidth]{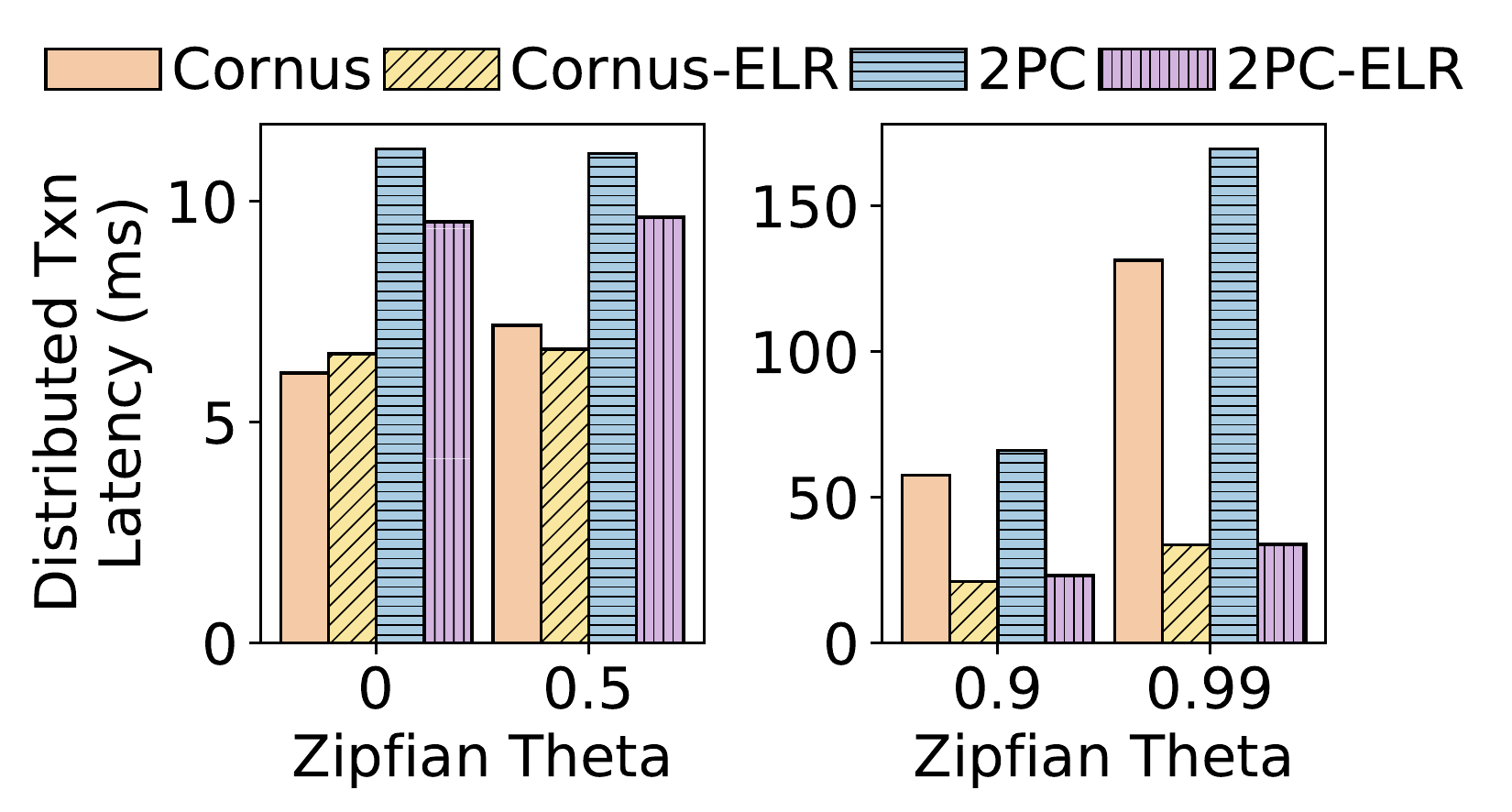}
        \vspace{-.2in}
        \caption{\edit{Latency (Redis)}}\label{fig:elr_a}
    \end{subfigure}%
	\begin{subfigure}[b]{0.48\linewidth}%
    	\center
        \includegraphics[width=\linewidth]{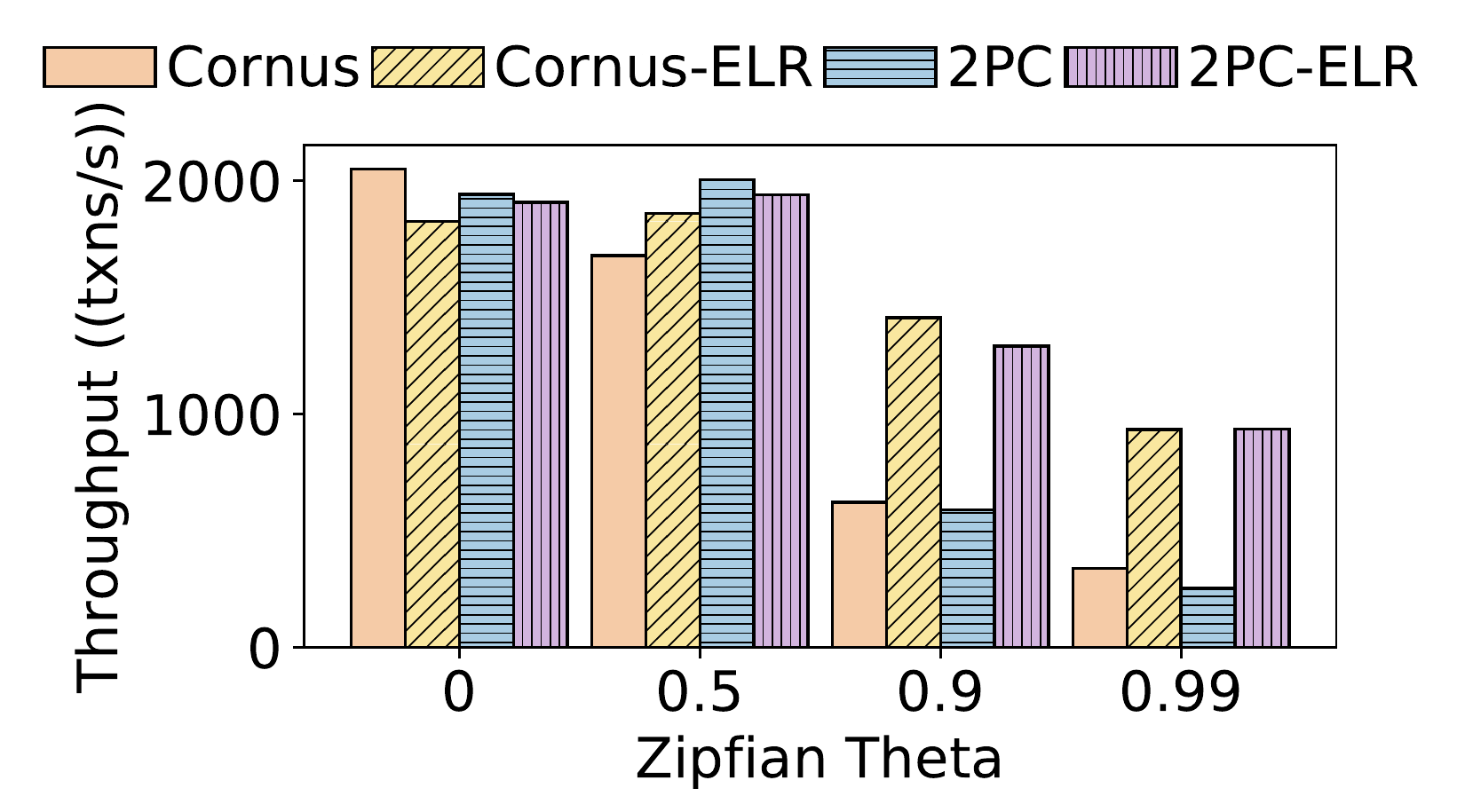}
        \vspace{-.2in}
        \caption{\edit{Throughput (Redis)}}
        \label{fig:elr_b}
    \end{subfigure}%
    \caption{\edit{Cornus and 2PC with speculative precommit}} 
    \label{fig:elr}
\end{figure}

\begin{figure}[t]
	\begin{subfigure}[b]{0.48\linewidth}%
    	\center
        \includegraphics[width=\linewidth]{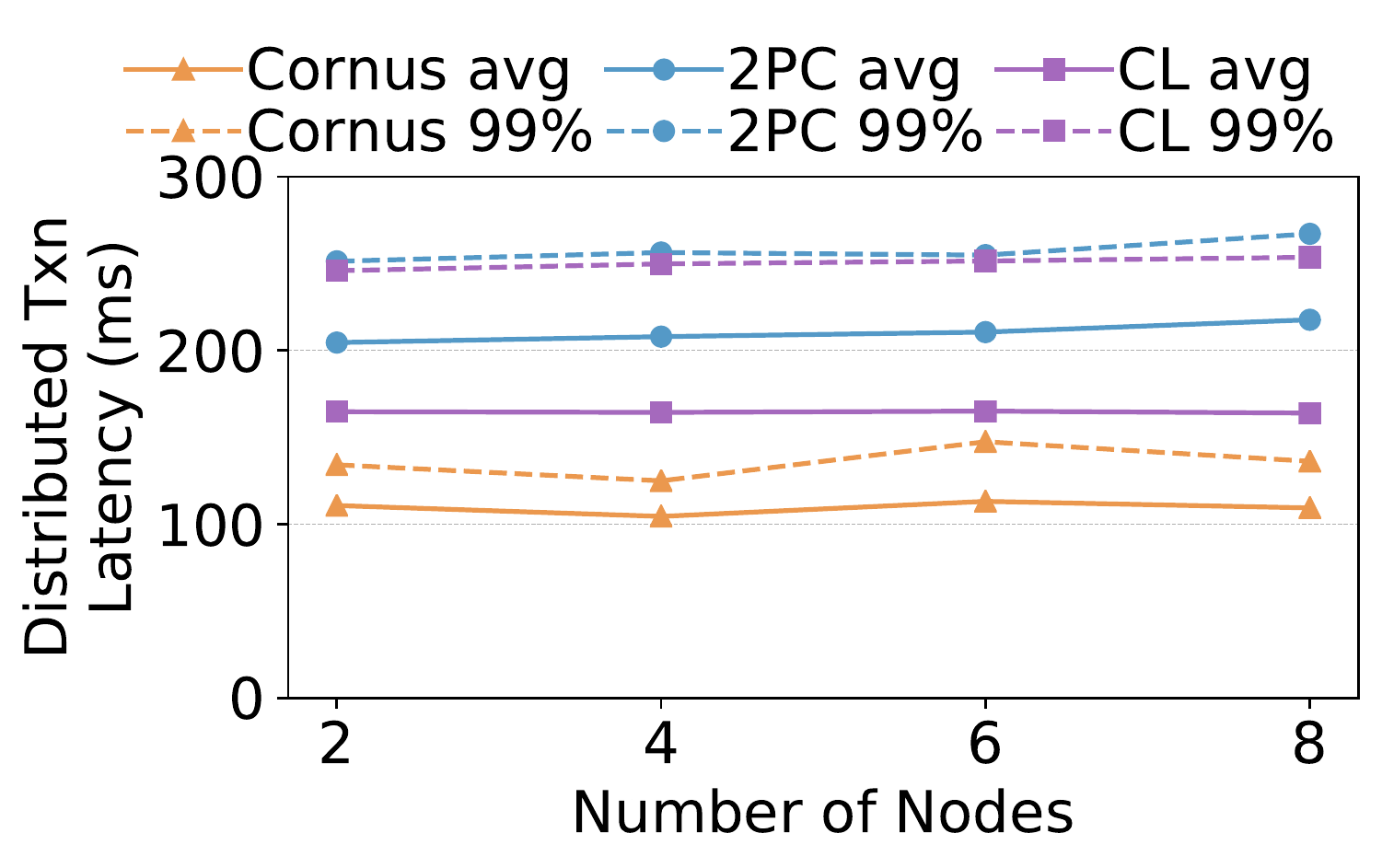}
        \vspace{-.2in}
        \caption{\edit{Latency (Redis)}}\label{fig:cl_a}
    \end{subfigure}%
	\begin{subfigure}[b]{0.48\linewidth}%
    	\center
        \includegraphics[width=\linewidth]{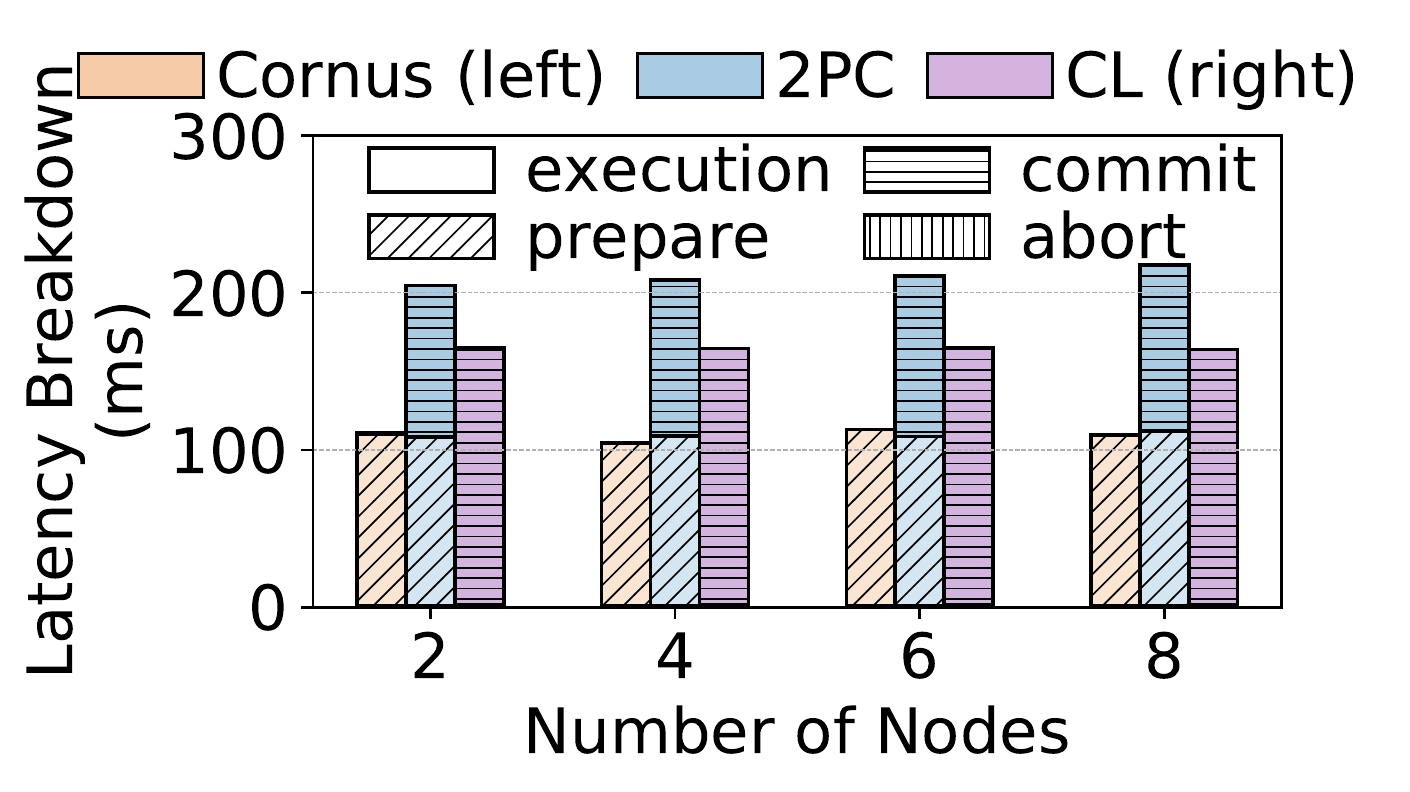}
        \vspace{-.2in}
        \caption{\edit{Latency (Redis)}}
        \label{fig:cl_b}
    \end{subfigure}%
    \caption{\edit{Coordinator log}}
    \label{fig:cl}
\end{figure}

\vspace{0.05 in}
\noindent \textbf{\edit{Speculative Precommit}}
\vspace{0.05 in}

\edit{The first optimization is to speculatively presume commit in the prepare phase. This optimization makes the assumption that a transaction entering the prepare phase is unlikely to abort due to system crashes. Therefore, a transaction can allow others to read its pre-committed data while waiting for the log to be persistent. Many previous papers~\cite{hstore, orleans-txn, graefe2013controlled, soisalon1995partial, kimura2012efficient} have studied this optimization.} 

\edit{This 2PC optimization is equally applicable to Cornus. We implement it in both Cornus (i.e., Cornus-ELR) and 2PC (i.e., 2PC-ELR) following a state-of-the-art protocol for distributed systems~\cite{guo2021lock}. Our implementation is based on pessimistic concurrency control, and thus we refer it as Early Lock Release (ELR) in this paper. \figref{fig:elr} compares Cornus and 2PC with and without the speculative precommit in the YCSB workload. In this experiment, we use Azure Redis as the remote storage and vary the contention levels on the x-axis.
As shown, ELR can significantly improve both 2PC and Cornus, particularly when workload contention is high. 
Specifically, Cornus and 2PC can improve by 175\% and 267\% with zipfian theta of 0.99 in throughput.
With the optimization, 
the difference between Cornus and 2PC in throughput and latency decreases under high contention since contention becomes the dominating factor as shown in~\figref{fig:redis_zipf_b}. 
In short, for systems that rarely crash during the prepare phase, this technique can be applied to Cornus to further improve the performance under high contention. 
}

\vspace{0.05 in}
\noindent \textbf{\edit{Coordinator Log}}
\vspace{0.05 in}

\edit{Another common 2PC optimization is to let the coordinator log on behalf of all participating nodes so that the transaction does not wait for other nodes to persist the logs. In our implementation, we ask the coordinator to log for all partitions during the prepare phase. At the same time, we send prepare requests to participants, which reply with their votes without logging. Upon receiving the votes, the coordinator makes decision and appends it to its log.}

\edit{We implement this technique following the ideas in previous works~\cite{stamos1990low, stamos1993coordinator} and compare it with 2PC and \name quantitatively. In this experiment, we use Azure Redis as the storage service, where the latency of writing to storage is around 443ms. In \figref{fig:cl}, we see that having the coordinator handle all logging (i.e., CL for Coordinator Log) outperforms the baseline 2PC by 33\% in terms of average latency since it batches all the logs into one write request to storage. The improvement is significant especially with such high latency of writing to Redis.
However, CL still underperforms Cornus by 50\% since the coordinator needs to wait until both the data and the decision are logged by coordinator, while Cornus directly replies to the user without waiting for the commit decision to be logged.}

\edit{Note that the coordinator log optimization has several limitations compared to 2PC or Cornus. First, it increases the size of acknowledgement messages. 
Second, it increases the complexity of recovery and raises security concerns. Specifically, it violates site autonomy which requires internal information concerning the local execution of transactions such as log records to remain private to a site and not be exported~\cite{abdallah1997non, abdallah1998one}. Although some works including \name allow other sites to access the log of transaction states, other sites will not access actual user data in \name (the second requirement discussed in~\secref{sec:deployment}). }

\vspace{.05in}
\noindent\textbf{\edit{Integration with Replication Protocol}}
\vspace{.05in}

\edit{
While \name 
directly runs on top of a highly-available storage service through a consensus-agnostic interface, many prior works~\cite{gray2006, kraska2013mdcc,taft2020cockroachdb,zhang2018building} manage the replication on their own and co-design 2PC with replication protocols for further optimizations. 
Although these protocols cannot be directly applied to existing storage services, we still evaluate them to show the potential optimization space when having different assumptions. 
}

\edit{We first performed theoretical evaluation on the expected latency for protocols combining with Paxos. Table~\ref{table:rtt} shows the number of Round Trip Times (RTTs) each protocol has on the critical path --- from the time the coordinator starts the protocol until the decision can be returned to the user. We use A + B = C to show (A) the number of RTTs in the prepare phase, (B) the number of RTTs in the commit phase, and (C) the sum. The table also shows the requirements.}

\edit{For 2PC and Cornus, we assume each process (coordinator/participant) runs an instance of Multi-Paxos in the underlying storage. When a participant sends a log request to the storage, it sends it to the leader of a Paxos instance, which refers to the proposer that has already run phase 1 to become a stable leader. 
The leader then initiates the second round of Paxos using one RTT and then gets back to the participant. Cornus can eliminate the coordinator logging from the critical path which corresponds to 2 RTTs (one for a participant communicating with the Paxos leader and one for the second round of Paxos).}

\edit{Cornus (optimization) refers to Cornus with optimization 1 discussed in  \secref{ssec:opt}. It can save the latency of logging acknowledgment from Paxos leader to participant (0.5 RTT) by forwarding the message to the coordinator. This optimization requires the storage to be able to send the message to an extra recipient but the replication details are still completely handled by the storage. }

\edit{
Cornus (co-location) and 2PC (co-location) represent the designs that co-locates a participant with the Paxos leader, i.e., the participant initiates the second round of Paxos by talking to all the replicas directly instead of asking the leader to initiate the process. Compared with naive Cornus and 2PC, this optimization can save the 1 RTT of Paxos leader communicating with other acceptors. However, this assumes that participants manage the replication process explicitly. }

\edit{Paxos Commit~\cite{gray2006} and MDCC~\cite{kraska2013mdcc} are two related works introduced in Section 2 and Section 6. MDCC-Classic refers to a protocol in the paper following the framework of Paxos Commit. These protocols apply all the optimizations discussed above. Specifically, during the prepare phase, a participant directly talks to all acceptors to log and all acceptors forward the logging acknowledgment to the coordinator. The coordinator then learns all the votes from the quorum. It can save 1 RTT for inter-replica communication and 0.5 RTT for logging acknowledgment sent from Paxos to coordinator compared with Cornus. However, it also has all the corresponding requirements including participant/coordinator coordinating replication and acceptors forwarding logging acknowledgement to the coordinator. }

\begin{figure}[t]
    \begin{subfigure}[b]{0.48\linewidth}%
    	\center
        \includegraphics[width=\linewidth]{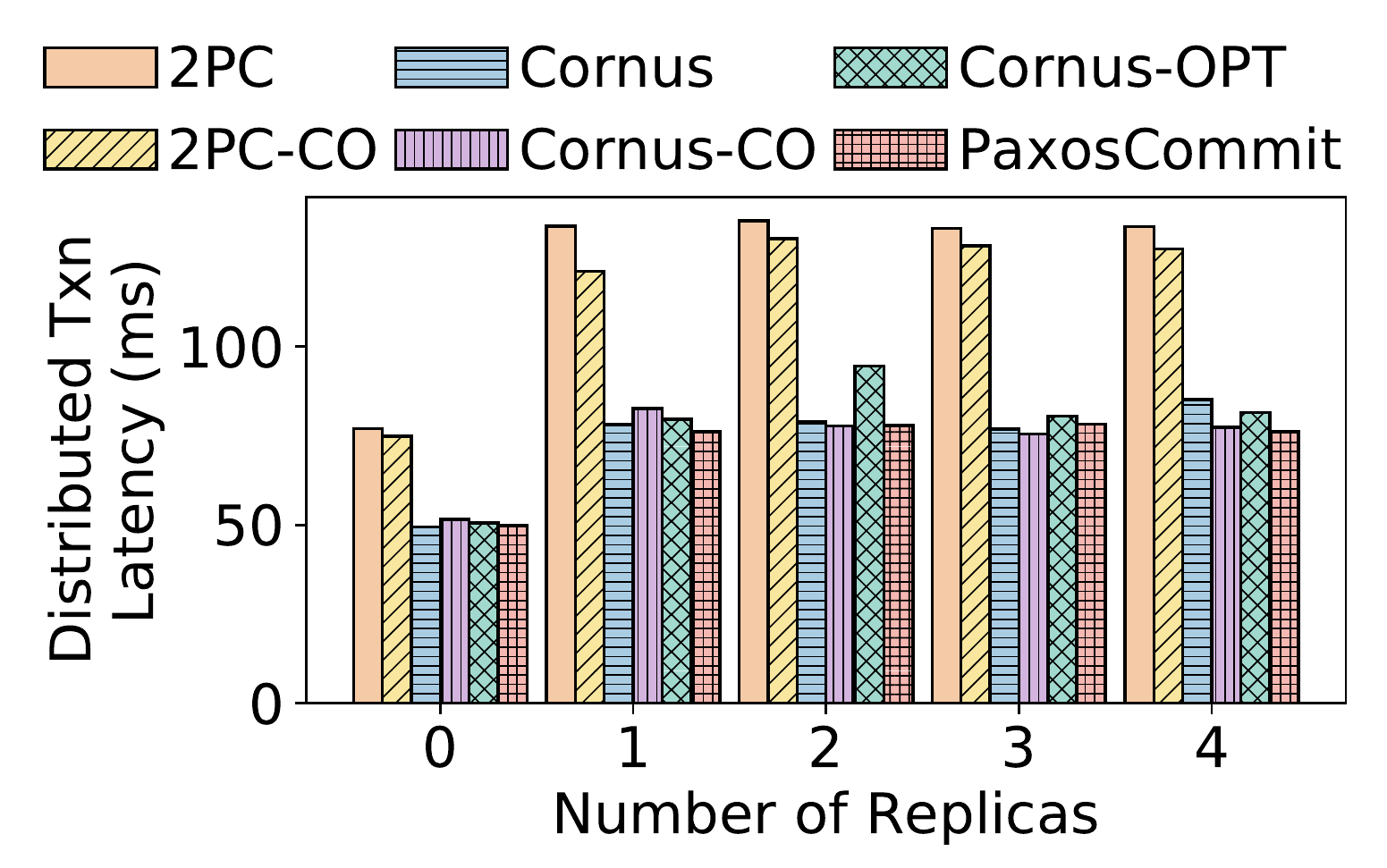}
        \caption{same region (RTT = 0.46 ms)}
        \label{fig:num_replica_same}
    \end{subfigure}
    \begin{subfigure}[b]{0.48\linewidth}%
    	\center
        \includegraphics[width=\linewidth]{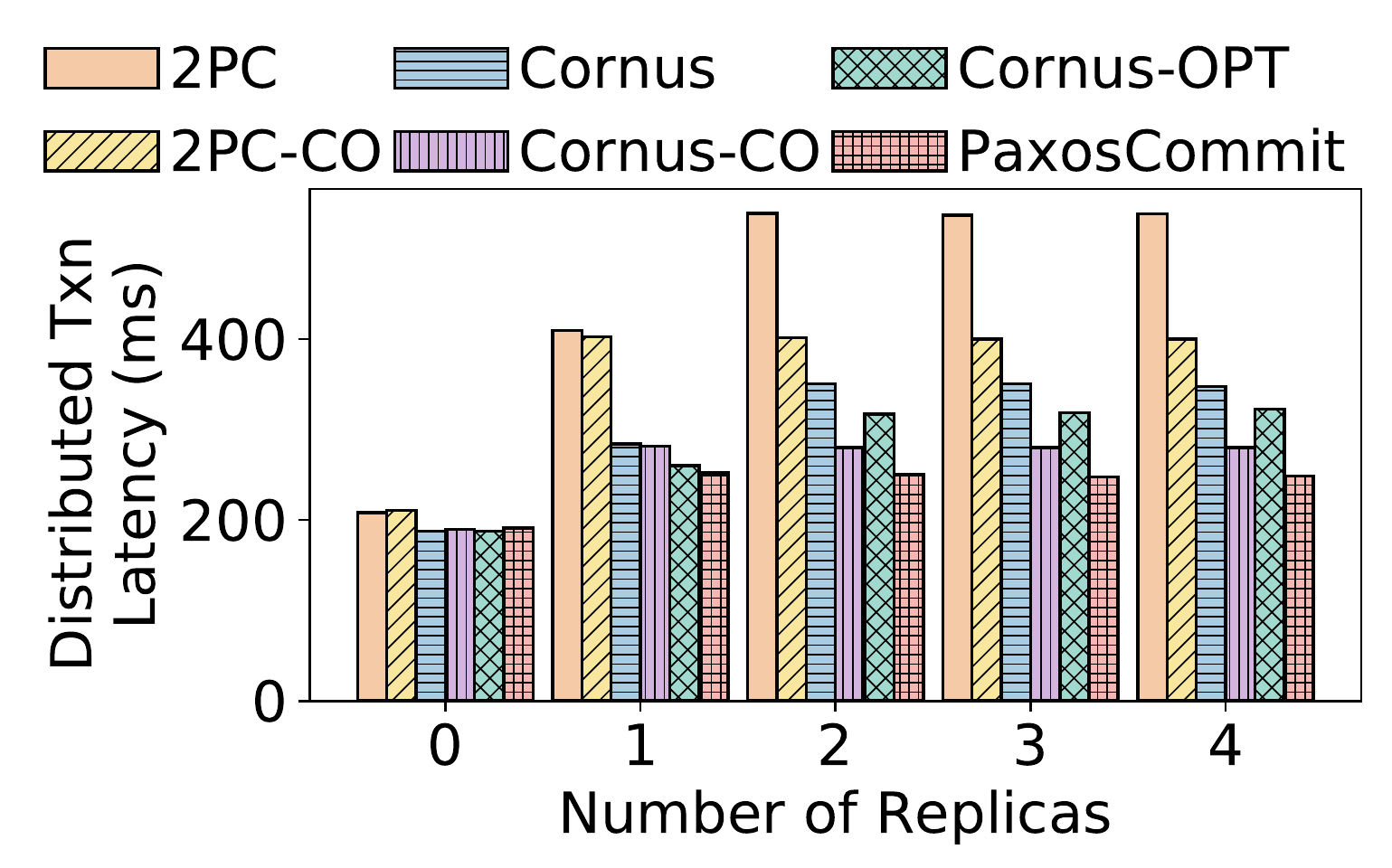}
        \caption{across region (RTT = 68 ms)}
        \label{fig:num_replica_cross}
    \end{subfigure}
    \caption{\edit{Cornus and 2PC with customized storage}}
    \label{fig:num_replica}
\end{figure}

 \edit{Besides the theoretical analysis, we also perform quantitative evaluation of these protocols with a self-implemented disaggregated storage and vary the number of replicas in the underlying storage. We run the experiment in two situations --- one with all replicas in the same region (\figref{fig:num_replica_same}) and one with replicas distributed across regions from US. East to US. West (\figref{fig:num_replica_cross}). Overall, the experiment results confirm the theoretical computation in Table~\ref{table:rtt} --- the relative performance maps to the expected differences in expected count of RTTs and the absolute improvement increases as RTT increases. The performance difference also increases with more replicas. This set of experiments demonstrate how much performance may be sacrificed when treating the storage as a black box with a consensus-agnostic abstraction, and the potential optimization space with co-designing 2PC with consensus to different degrees.
 }

\begin{table}
\centering
\begin{tabularx}{\linewidth}{p{0.65in}p{0.6in}p{1.7in}}
  \hline
  Protocol & \# RTT & Extra Requirements\\ 
  \hline
  2PC & 3 + 2 = 5 & - \\
  \hline
  Cornus & 3 + 0 = 3 & Storage supports conditional write \\
  \hline
  Cornus (optmization) & 2.5 + 0 = 2.5 & Leader of Paxos can forward a message to coordinator\\
  \hline
  2PC (co-location) & 2 + 1 = 3 & Participant coordinates replication \\
  \hline
  Cornus (co-location) & 2 + 0 = 2 & Participant coordinates replication \\
  \hline
  Paxos Commit / MDCC-Classic & 1.5 + 0 = 1.5 & Participant coordinates replication; Acceptors forward messages to coordinator to learn from quorum\\
  \hline
\end{tabularx}
\caption{Time complexity for protocols integrating with Paxos or its variations}
\label{table:rtt}
\end{table}

\section{Related Work} \label{sec:related_work}
This section describes related works on optimizing 2PC. \edit{We categorize prior works into three categories: techniques reducing latency, techniques addressing blocking, and codesign of 2PC and replication to address both problems.}

\subsection{\edit{Techniques Reducing Latency}}

\vspace{.05in}
\noindent\textbf{\edit{Centralized Logging. }}
\edit{
Centralized logging 
asks the coordinator to log for all participants on behalf of them to reduce latency. A line of previous work follows this idea including \textit{coordinator log}~\cite{stamos1990low, stamos1993coordinator}, \textit{implicit yes vote}~\cite{al1995two}, and Lee and Yeom's protocol~\cite{lee2002single}.}
The main idea is to have each participant send its log with its acknowledgment of its prepare request to the coordinator. The coordinator force writes those acknowledgments to its log along with its commit decision so that it eliminates the logging latency in prepare phase. However, as discussed in Section 5.6, these designs increase the size of acknowledgment, increase the complexity of recovery, and raise security concerns as they violate site autonomy. 

\vspace{.05in}
\noindent\textbf{\edit{Early Prepare During Execution.}}
\edit{Another technique to reduce 2PC latency is to let participants prepare during execution so that a transaction can skip the prepare phase at commit time. Many previous works incorporate this idea to reduce the commit latency~\cite{stamos1990low, stamos1993coordinator, al1995two, lee2002single}. 
For example, in \textit{Early Prepare} (EP)~\cite{stamos1990low}, each participant forces a prepare record before replying to the coordinator on receiving a work request during execution. It also requires the coordinator to record the identity of the participant in the log before sending the work request. 
However, if the transaction accesses more than one partition or the work requests for each partition cannot be batched into one request, EP needs to pay more logging in the critical path than 2PC. }

\edit{To address this issue, subsequent works try to combine this technique with centralized logging or decentralized decision.} For example, implicit yes vote~\cite{al1995two} applies both centralized log and early prepare. It asks participant to piggyback its log in the acknowledgement of every work request and treats the acknowledgement as an implicit yes vote. However, it still suffers from the limitations of centralized log discussed above and additionally restrict the design of concurrency control. For example, protocols like optimistic concurrency control and two-phase locking with wound-wait~\cite{bernstein1987concurrency} need to be carefully redesigned since a participant may abort after sending the acknowledgement at the end of execution phase~\cite{abdallah1998one}.

\vspace{.05in}
\noindent\textbf{\edit{Speculative Pre-Commit. }}
\edit{If a transaction that has entered the prepare phase is unlikely to abort due to system crashes, the database can speculatively presume commit in the prepare phase. 
Specifically, a transaction can let others read its pre-committed data while waiting for the log to be persistent. Many previous works~\cite{hstore, orleans-txn, graefe2013controlled, soisalon1995partial, kimura2012efficient} have studied this optimization known as early lock release~\cite{kimura2012efficient} and controlled lock violation~\cite{graefe2013controlled}. In \secref{ssec:eval-optimization}, we have shown this optimization can be applied to both 2PC and \name leading to similar throughput improvement and latency reduction. 
}

\balance

\subsection{\edit{Techniques Addressing Blocking}}
\vspace{.05in}
\noindent\textbf{\edit{Extra Network Roundtrip.}}
Skeen~\cite{skeen1981nonblocking} gave necessary and sufficient conditions for a correct non-blocking commit protocol, known as \emph{the fundamental nonblocking theorem}. 
It proved that a fifth state in addition to initial, wait, abort, and commit can be added to avoid blocking with the same assumptions made in 2PC. 
Adding this state requires one more network round trip, resulting in a three-phase commit (3PC) protocol. Although it solves the blocking issue, 3PC magnifies the problem of latency delay in 2PC. 

\vspace{.05in}
\noindent\textbf{\edit{Extra Message Count.}}
\edit{Some protocols decrease the chance of blocking by asking each site to broadcast the messages upon receiving. } EasyCommit~\cite{gupta2018easycommit} solves the blocking problem by requiring each participant to forward the decision to other participants before logging it.
However, the protocol satisfies the atomic commit properties only if at least one participant receives the forwarded message  before the sender flushes the decision to its log. It also introduces extra messages during normal execution and increases the complexity when failure happens. 
Babaoglu and Toueg~\cite{babaoglu1993understanding} propose a non-blocking atomic commit protocol based on 2PC. It applies three strategies: (1) synchronizing clocks on different sites so that out-of-time messages can be ignored; (2) having participants forward the decision to other participants upon receiving the message from the coordinator; and (3) presuming abort instead of running a termination protocol upon timeout. The first two strategies enable the last to avoid blocking. However, the protocol introduces more communication in the absence of failure, and
the algorithm relies on synchronized clocks, a non-trivial requirement for real-world applications.

\subsection{Co-design of 2PC and Replication} 
Many prior works optimize 2PC through the co-design of 2PC and replication protocols (e.g., Paxos) to solve latency and blocking problems at the time.  
Gray and Lamport~\cite{gray2006} proposed Paxos Commit, which is a theoretical framework for optimizing 2PC and Paxos with a space of optimizations. It proposed some Paxos-specific optimizations such as pre-preparing acceptors and piggybacking messages of 2PC and Paxos. 
Several implementations followed the spirit of Paxos Commit and made adaption for their scenarios~\cite{kraska2013mdcc,zhang2018building}.

Kraska et al.~\cite{kraska2013mdcc} introduced Multi Data Center Consistency (MDCC). It assumes resource manager and Paxos leader are co-located on the same site so that it can piggyback the messages of Paxos with 2PC requests to save message roundtrips. They also propose a leaderless version combining 2PC and Fast Paxos. This version assumes that each acceptor can perform conflict detection for optimistic concurrency control independently and produce the same validation result. It can further reduce latency when conflicts are rare. Moreover, it optimizes for commutative operations by using update intents ("options") instead of the actual updates.

A recent work of parallel commit protocol~\cite{vanBenschoten2019, parallelcommits} in CockroachDB (CRDB)~\cite{taft2020cockroachdb} uses primary copy replication based on Raft to determine whether the commit operation is completed. It co-locates the leader of Raft with participants (resource manager) while Cornus elevates this check to an external wrapper over cloud storage, whose replication algorithm is a black box.

Zhang et al.~\cite{zhang2018building} introduced TAPIR. It used a customized replication protocol, Inconsistent Replication, to relax consistency in storage replicas and relies on application protocols to resolve inconsistencies. 
Mahmoud et al.~\cite{mahmoud2013low} proposed a Replicated Commit protocol that runs 2PC at different data centers. It uses Paxos to reach consensus across data centers to determine if a transaction should commit.   
Yan et al.'s Carousel~\cite{yan2018carousel} runs 2PC in parallel with the execution phase and state replication, but assumes the read and write key sets are known in advance. 
Fan et al.~\cite{fan2019ocean} combines concurrency control, transaction commit, and replication in a single protocol to better utilize data locality and reduce cross-data center networks. 

\edit{Finally, deterministic databases~\cite{thomson12,faleiro2017high,faleiro15,qadah2020q,lu2020aria,thomson2010case} take a drastically different approach to handle 2PC and replication. 
Instead of treating computation nodes and storage service as horizontally separated layers, a deterministic database vertically partitions the cluster into replicas, and ensures consistency across replicas by running the same input transactions deterministically in all the replicas such that they produce identical results. 
Deterministic databases also simplify 2PC since only the inputs of transactions are made persistent and no logging happens during transaction execution. Compared to 2PC and \name, deterministic databases have several limitations. For example, transactions need to be one-shot (i.e., cannot support multiple interactions with the DBMS); transactions must run in batches such that a single long-running transaction prolongs response time for the entire batch; most deterministic databases (except Aria~\cite{lu2020aria}) require knowing the read/write sets of transactions before execution.
}
\section{Conclusion} \label{sec:Conclusion}
We proposed \name, a protocol optimizing 2PC for storage disaggregation, an architecture widely used by modern cloud databases. \name solves both the \textit{long latency} and the \textit{blocking problem} in 2PC by leveraging the new features provided by the architecture. We formally proved the correctness of \name and evaluated it on top of practical storage services including Redis and Azure Blob Storage. Our evaluations on YCSB show a speedup of up to 1.9$\times$ in latency.

\bibliographystyle{ACM-Reference-Format}
\bibliography{main.bib,db.bib}

\end{document}